\documentclass[prd,preprintnumbers,nofootinbib,superscriptaddress,onecolumn,notitlepage]
{revtex4}

\usepackage{cancel}
\usepackage{amsmath,mathrsfs,amscd,graphicx}
\usepackage{mciteplus}
\usepackage{multirow}
\usepackage{color}
\usepackage{cancel}
\usepackage{rotating}
\usepackage{slashed}
\usepackage{subfig}
\usepackage{textcomp}
\usepackage[final]{pdfpages}
\usepackage{array}
\usepackage{float}
\usepackage{url}
\usepackage{textcomp}
\usepackage{amsfonts, latexsym, epsfig}
\usepackage{bm}
\usepackage{times}
\usepackage{epsfig}
\usepackage{amssymb}
\usepackage{tikz}
\usepackage{slashed}
\usepackage{hyperref}
\usepackage{verbatim}
\usepackage[normalem]{ulem}
\usepackage[utf8]{inputenc}
\usepackage{diagbox}
\usepackage{titlesec}
\def\beq{\begin{equation}}
\def\eeq{\end{equation}}
\def\be{\begin{equation}}
\def\ee{\end{equation}}
\def\bea{\begin{eqnarray}}
\def\eea{\end{eqnarray}}

\definecolor{dpmagenta}{rgb}{0.8, 0.0, 0.8}

\allowdisplaybreaks[3]
\begin{document}
\title{Electroweak corrections to Higgs boson production via W W fusion at the future LHeC}
\author{Bowen Wang~}
\email{bowenw@hznu.edu.cn}
\affiliation{Zhejiang Institute of Modern Physics and School of Physics, Zhejiang University, Hangzhou, Zhejiang 310027, China}
\affiliation{School of Physics, Hangzhou Normal University, Hangzhou, Zhejiang 311121, China}
\author{Kai Wang~}
\email{wangkai1@zju.edu.cn}
\affiliation{Zhejiang Institute of Modern Physics and School of Physics, Zhejiang University, Hangzhou, Zhejiang 310027, China}
\author{Hanying Xiong~}
\email{21736003@zju.edu.cn}
\affiliation{Zhejiang Institute of Modern Physics and School of Physics, Zhejiang University, Hangzhou, Zhejiang 310027, China}
%

\begin{abstract}
    Precision measurement of quark Yukawa couplings is a crucial aspect of Higgs property study. 
	Proposed as a future upgrade of the Large Hadron Collider (LHC), the Large Hadron electron Collider (LHeC) 
	provides opportunities to probe quark Yukawa couplings with a high precision because of relatively low rate 
	from the QCD background as compared with that of the Higgs processes at the LHC. For this purpose, it is important 
	to have a precision prediction of the Higgs production rate at the LHeC. The leading production channel of the Higgs boson 
	at the LHeC is via weak boson fusion (WBF) which has the unique kinematic feature of forward tagging jets. 
	As QCD corrections are suppressed by the simple color structure of the process, electroweak radiative corrections become particularly important. 
	In this paper, we focus on the electroweak radiative corrections at next-to-leading order for the Higgs-boson production via charge current WBF
    processes at the LHeC. In the two renormalization schemes we use, the loop corrections are respectively at the level of 
	$9\%$ and $18\%$ relative to the leading order result, for a center-of-mass energy at $1.98$ TeV, and the next-to-leading order 
	results in both schemes agree up to a truncation error expected from the perturbative method. 
	The size of the EW corrections exceeds that from QCD radiations and is shown to be important for the study of Higgs phenomenology.
	
\end{abstract}

\maketitle

\section{Introduction}
Enormous efforts have been made to study the properties of the 125~GeV Higgs boson at an increasing precision level
since its discovery~\cite{Aad:2012tfa,Chatrchyan:2012xdj}. In particular, the measurement of Higgs couplings to 
various particles~\cite{Cadamuro:2019tcf,Langford:2021osp,Ordek:2022ppj} is an important part of the endeavor. 
Recent experiments have achieved a percent level accuracy in the extraction of Higgs couplings to  gauge bosons 
and the top quark~\cite{ATLAS:2020qdt,CMS:2020gsy}, which are all in agreement with the standard model (SM) predictions.
These measurements are preferably carried out in the leptonic decay channels of the Higg boson. For instance, 
in Ref.~\cite{CMS:2020gsy} the Higgs coupling to the $Z$ boson is extracted from $H \rightarrow ZZ^* \rightarrow 4\, \text{leptons}$;  
the top quark Yukawa is measured via the Higgs production associated with a top quark pair, with $H\rightarrow \text{leptons}$.

In contrast, the hadronic final states from Higgs decays are much more difficult to probe because of the large rate 
for the background jets (produced not via intermediate Higgs bosons).
The $H\rightarrow \tau^+\tau^-$ with subsequent hadronic decays produces narrow $\tau$ jets, 
but the jet shape alone is insufficient for identifying the process from the hadronic background.
Nonetheless, constraining of $H$-$\tau$ Yukawa coupling can be carried out in a production mode commonly referred to as the 
``weak boson fusion'' (WBF), where a pair of intermediate W or Z bosons is radiated by the colliding fermions and couples to a Higgs. 
The energetic incoming particles evolve to a final state with a pair of fermions (at the parton level)  separated by a large rapidity gap, 
in association with the Higgs and other particles produced in the central rapidity region. 
The transverse momenta of the two fermions are of order the weak boson mass. 
This configuration of event is successfully utilized for suppressing backgrounds in the search of 
$H \rightarrow \tau\tau$ at the Large Hadron Collider (LHC), for both hadronic and leptonic decays of $\tau$'s~\cite{CMS:2017zyp,ATLAS:2018ynr}.

The identification of $H$-$b$ and $H$-$c$ couplings via WBF at hardron colliders gets even more challenging~\cite{CMS:2015ebl,ATLAS:2016mzy} 
 because of the huge QCD multi-jet background.  On the other hand,  the Large Hadron electron Collider (LHeC)~\cite{LHeCStudyGroup:2012zhm} 
  is proposed partly to measure Higgs couplings such as $H$-$b$ and $H$-$c$. 
In this upgrade option for the LHC, a beam of electrons will be aligned to collide with the 7 TeV protons.
The electron beam energy is considered to range in 50-200 GeV, making the facility a factory to produce Higgs bosons mainly via WBF.
   Studies have shown the prospect of making the measurement of Yukawa couplings at the LHeC, 
  based on the analysis using leading order (LO) matrix elements in the simulations~\cite{Han:2009pe,LHeCStudyGroup:2012zhm,LHeC:2020van,Li:2019xwd}. 
  An account for the next-to-leading order (NLO) corrections is needed both for a precise knowledge of
  the signal cross section and for determination of various selection criterions affected by the distortion of distributions at NLO.

 Computation of NLO QCD and (part of) QED corrections for WBF production of the Higgs has been performed by  
 Blumlein~\cite{Blumlein:1992eh}, and Jager~\cite{Jager:2010zm}. 
 The QCD corrections to the total cross section are shown to be at the level of a few percent, while their impact on differential distributions
 is found to be more sizable, i.e. as large as $10\%\sim 20\%$. The dominant QED radiative corrections
 are about -5\%. As we will show, it is important to include all the NLO EW processes, 
 which give rise to substantial corrections in both integrated cross section and differential distributions.
 
  We note that part of the EW loop effect has been calculated in a study~\cite{Li:2019jba} to estimate the significance of the triple Higgs-self
 coupling at the LHeC. However, the calculation only involves a small subset of the loop diagrams associated with the coupling.
 We shall see that the background (with no Higgs self coupling) cross section in this study can be reduced by including all
 EW loop processes, which makes a negative contribution and therefore yields a better constraint on the Higgs self coupling.

 In this paper we present a calculation of the full NLO EW corrections to the WBF cross section in $e$-$p$ collisions.  To make the discussion
 general we consider only the partonic scattering amplitude with an on-shell Higgs, and do not simulate the final state to which the Higgs decays.
 For simplicity, the focus will be on the charge current(CC) WBF, which is dominant over the neutral current(NC)
 process.\footnote{The integrated cross section of CC WBF is about 5 times that of NC WBF at the energy regime of the LHeC.} 
 Moreover, the outgoing electron of the NC process could be tagged in a broad kinematical region where it could be distinguished from the CC process.

The past two decades have seen the fast development of automation programs for the calculation of QCD and EW corrections for various scattering 
processes at one loop accuracy~\cite{Alwall:2014hca,Buccioni:2019sur,Sherpa:2019gpd,Bellm:2015jjp,Alioli:2010xd,Alioli:2012fc,Kilian:2007gr}. 
However, these tools prioritize the processes in $p$-$p$ and $e^+$-$e^-$ scatterings, and to our best knowledge, 
no implementation is made as yet in the publicly available Monte-Carlo programs to fully automatize the NLO calculation for e-p collisions. 
To treat the WBF process at the LHeC, we need to organize different parts of the calculation on our own.
The calculation proceeds in the dipole subtraction formalism~\cite{Catani:1996vz}, first designed for a unified treatment of various singularities encountered
 in scattering amplitudes with real and virtual QCD radiations. The method was then adapted to the calculation of QED radiative corrections
 and generalized to include the radiative effect that is sensitive to the mass of the initial or final fermions~\cite{Dittmaier:1999mb,Catani:2002hc,Schonherr:2017qcj}.
We shall show how to consistently include the fermion mass effect in our calculation of the WBF process using the dipole method.

The rest of the paper is organized as follows. In Sec.~\ref{higgs_xsecs}, we give the description of the details for the analytical
and numerical parts of the calculation. In Sec.~\ref{pheno}, we discuss the numerical significance of the NLO EW corrections in both
total and differential cross sections of the WBF process. Then we conclude the paper in Sec.~\ref{conclusion}.

\section{Detailed calculation of the process}
\label{higgs_xsecs}
 
 \subsection{LO contribution}
\label{LO}
\begin{figure}[h]
\centering
    \includegraphics[scale=0.5]{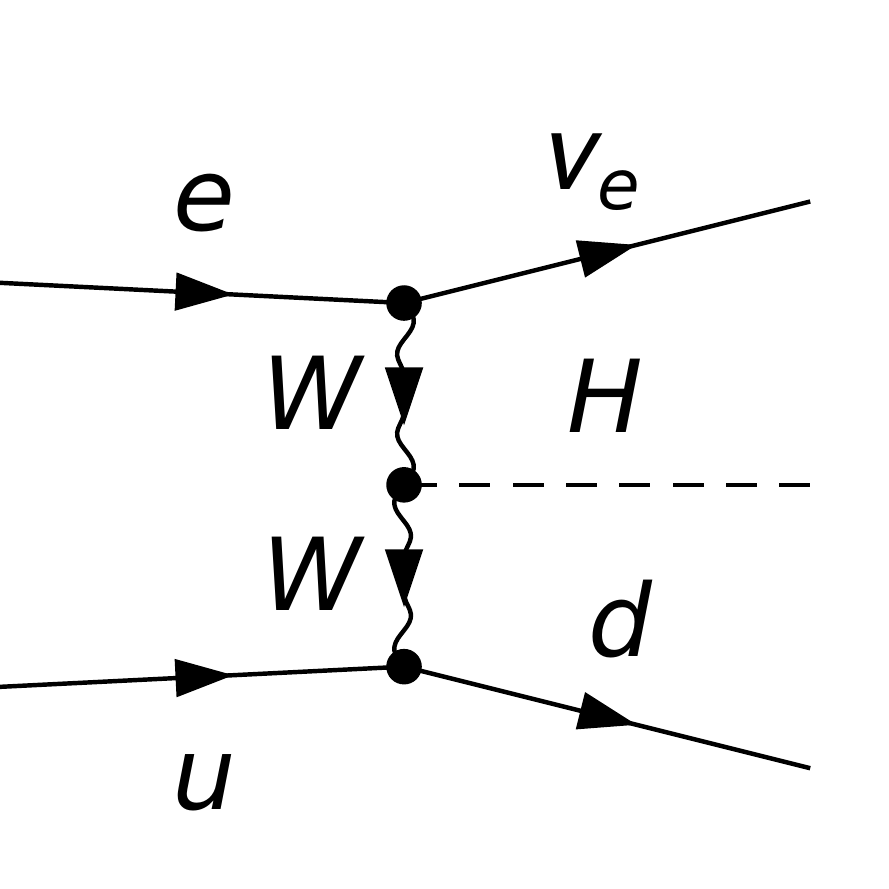} \qquad
	\caption{A representative leading order diagram for CC WBF at the LHeC.}
\label{born}
\end{figure}

The CC WBF $ep \rightarrow \nu_e jH$ is depicted at the leading order by the
partonic process $eq \rightarrow \nu_e q^{\prime} H$. An example diagram is shown in Fig.~\ref{born}.
 The cross section is given by
\begin{equation}
\begin{aligned}
	\sigma^{LO}=\sum_{q}\int_0^1 d\eta_q f_q(\eta_q,\mu_F) \frac{1}{2\hat{s}}d\Phi_3 B(P_A,p_q),    
\label{eq:LOxsec}
\end{aligned}
\end{equation}
where $B(P_A,p_q)$ denotes the born amplitude squared at the parton level with the incoming electron and quark momenta $P_A$ and $p_q$. 
The partonic center-of-mass (CM) energy squared is given by $\hat{s}=(P_A+p_q)^2$, and $d\Phi_3$ is the phase space element of the 3-body final state.
In this paper we assume no particular polarization for external particles, so 
that in $B(P_A,p_q)$ and all other squared amplitudes below, we sum and average over colors and spins of the initial and final states.
$\eta_q$ is the fraction of the proton momentum $P_B$ carried by the incoming quark $q$. The hadronic cross section is factorized at the scale $\mu_F$
into a quark distribution function $f_q(\eta_b,\mu_F)$, convolved with a squared hard scattering amplitude depending implicitly on $\eta_q$.
The evolution of $f_q(\eta_b,\mu_F)$ introduces NLO QCD and QED terms to the LO cross section, which is not a problem as long as we
do not double count these terms in our NLO calculation.
The dominant contribution comes from
the initial quark flavors $u$, $c$, $\bar{d}$ and $\bar{s}$, while the effect of $\bar{b}$ and $t$ is marginal and neglected.

\subsection{NLO EW corrections}
\label{NLO}
\begin{figure}[h]
\centering
    \includegraphics[scale=0.5]{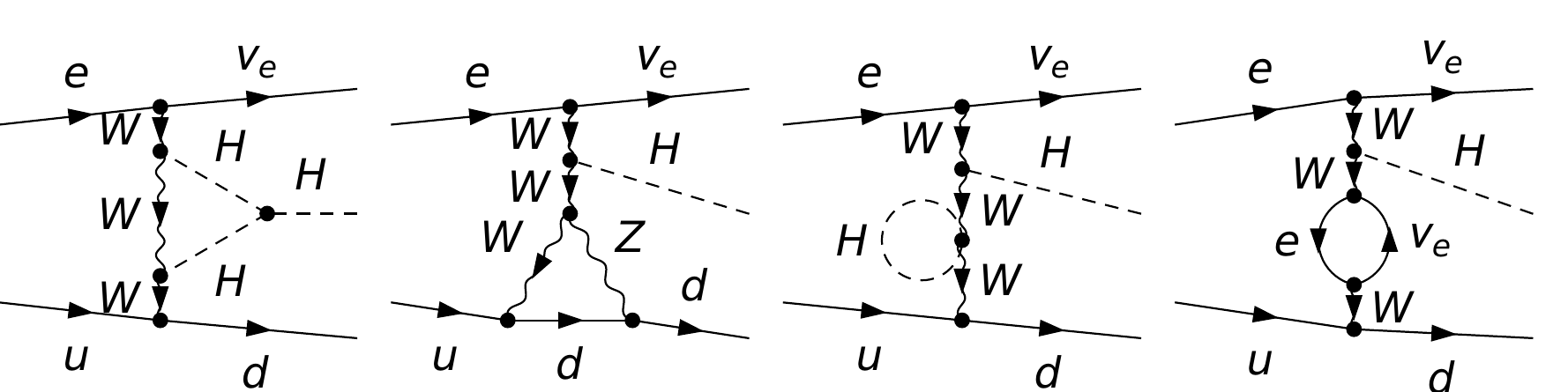}\qquad
    \includegraphics[scale=0.5]{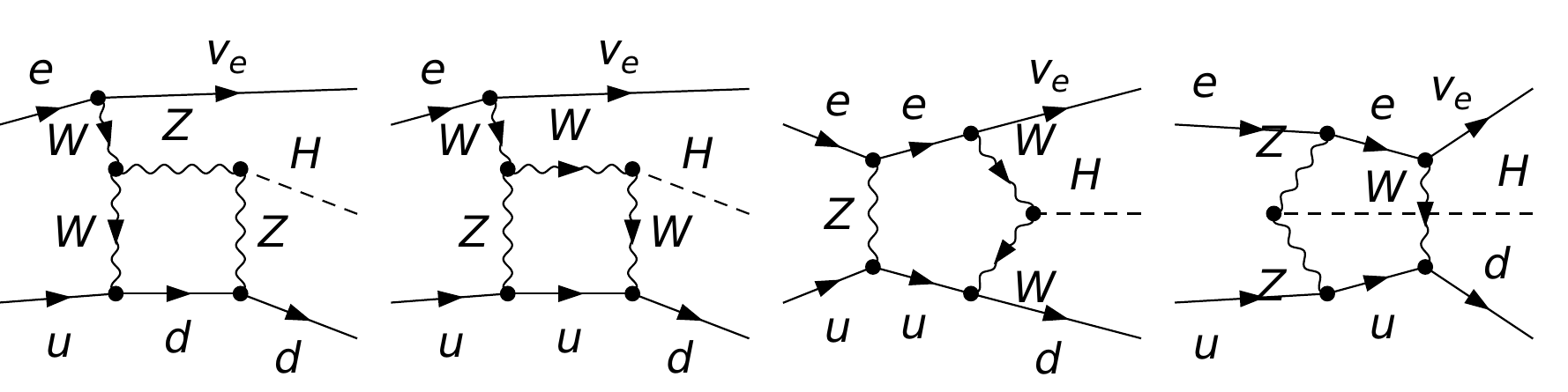}\qquad
\caption{Representative 1-loop diagrams for CC WBF at the LHeC with no soft/collinear divergences.}
\label{loop_NO_IR}
\end{figure}

\begin{figure}[h]
\centering
    \includegraphics[scale=0.5]{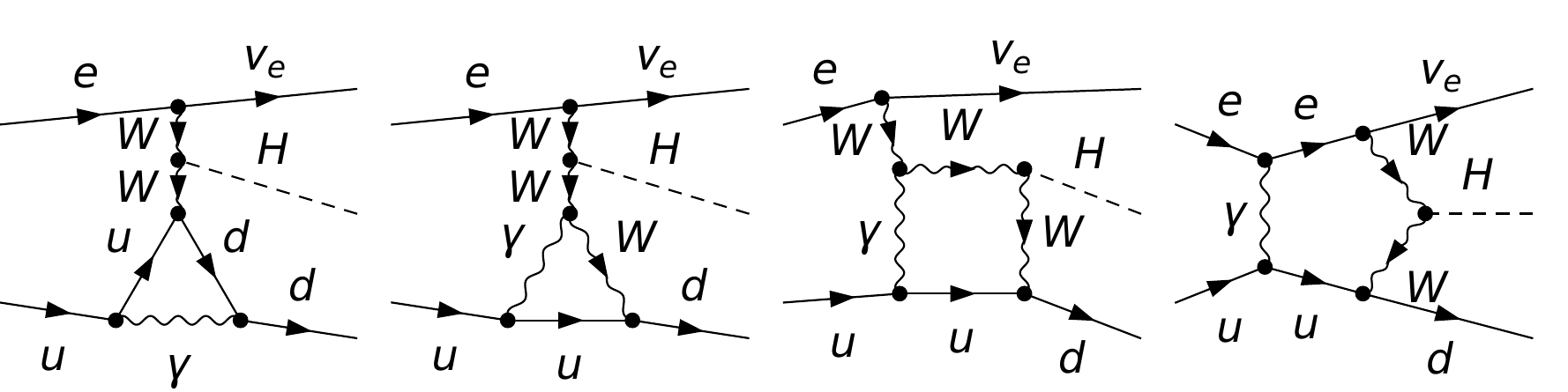}\qquad
\caption{Representative 1-loop diagrams for CC WBF at the LHeC with soft/collinear divergences.}
\label{ir_div}
\end{figure}

The numerically dominant NLO EW corrections are from the loop diagrams of various topologies,
ranging from self energy graphs to pentagons.
Soft and collinear singularities may arise in some of these diagrams when photons 
are present in the loop. Representative loop diagrams with and without these singularities are shown in 
Fig.~\ref{ir_div} and Fig.~\ref{loop_NO_IR}, respectively.

A consistent treatment of the singularities requires the inclusion of all diagrams with real emission of photons
, whose momenta can potentially be either soft or collinear to the emittee. In properly defined observables,
cancellation of part of the singularities takes place between real-emission and loop diagrams.
Mass singularities associated with initial state radiations are removed by the collinear
counter terms provided by the factorization procedure, to be discussed below. There are also real-emission processes induced by photons
that are treated as constituents of the initial state protons. In this case, no corresponding loop diagrams are present
and the singular terms are cancelled solely by the collinear counter terms.

Within the energy range of the LHeC, we treat $u$, $d$, $s$, and $c$ (and their anti-particles) as massless flavors, 
whose couplings with the Higgs are also neglected.
The contribution from $b$ and $t$ (and their anti-particles) is marginal and not included.
The role of the electron mass is somewhat intricate and will be discussed later.
With these simplifications at NLO, there are in total 628 loop diagrams produced by the program {\sc MadGraph5\_aMC@NLO}~\cite{Alwall:2014hca},
and 42 real emission diagrams by {\sc FeynArts}~\cite{Hahn:2000kx} with quark, anti-quark, and photon initial states.

\subsubsection{Renormalization}
\label{Ren}

The calculation of one-loop diagrams has been performed in the t'Hooft-Feynman gauge with the program {\sc MadLoop}~\cite{hirschi:2011pa}.
Several packages are linked by {\sc MadLoop} for the reduction of tensor integrals,
each based on one of two distinct procedures: the Tensor Integral Reduction~\cite{Passarino:1978jh,Davydychev:1991va} 
and the Ossola-Papadopoulos-Pittau method~\cite{Ossola:2006us}.
For simplicity, we neglect quark mixing and do not renormalize the quark mixing matrix. 

The  mass and field strength renormalization constants are determined within
the on-shell renormalization scheme, except for the field strength $Z_{AA}$ of the photon
that is closely related to the renormalization of the electric charge.
It can be obtained by imposing a renormalization condition at some momentum scale, or equivalently,
by specifying the form of the renormalized fine structure constant. In this work we adopt two schemes. The $G_{\mu}$ scheme is
defined by the choice~\cite{Denner:2019vbn}
\begin{equation}
\begin{aligned}
    \alpha_{G_{\mu}}=\frac{\sqrt{2}G_{\mu}M_W^2}{\pi}\left (1-\frac{M_W^2}{M_Z^2} \right )\approx \frac{1}{132},
\end{aligned}
\end{equation}
where the Fermi constant $G_{\mu}$ is determined from the muon decay experiment, 
while the $\alpha(M_Z)$ scheme is obtained by evolving the physical fine structure constant $\alpha(0)\approx 1/137$ 
from zero momentum transfer to the scale $M_Z$, yielding $\alpha(M_Z) \approx 1/129$.
The numerical value of $\alpha$ in both schemes differ from $\alpha(0)$, which is obtained by
requiring that all high order corrections to the electron-photon 3-point function vanish in the Thomson limit~\cite{Denner:1991kt,Denner:2019vbn}.
The differences are sensitive to the light fermion masses on which
 the loop contribution to the photon wave function depend. The sensitivity manifests as logarithmic terms
 of the form $\ln m_f$~\cite{Ellis:2007qk,Denner:2019vbn}, indicating a collinear divergence in the on-shell photon
 wave function as the fermion masses $m_f$ approach 0.
 It is straightforward to show that in the $G_{\mu}$ and $\alpha(M_Z)$ schemes, $Z_{AA}$, as well as the charge renormalization constant $Z_e$,
 are free from the collinear divergence, such that the definition of the
 physical charge on-shell is maintained. As long as no external photons are present in the LO
 process, we are able to consistently implement the renormalization constants and neglect the small mass $m_f$ in the calculation
 of the loop diagrams with dimensional regularization. 
 
 \subsubsection{Factorization}
\label{Fac}

After factorizing the parton distribution functions (PDFs) of the incoming quarks, their mass can be
safely set to zero in the partonic scattering amplitudes initiated by these flavors. The mass singularities from the initial state photon-quark
splitting signal our ignorance of the nonperturbative interactions of the partons in the proton below some factorization scale $\mu_F$, on which
the QED evolution of the PDFs depends. The corresponding poles in
the hard scattering process are subtracted by adding a collinear counter term to the squared amplitudes. In this calculation we use the PDF set
{\tt CT14qed\_inc\_proton}~\cite{Schmidt:2015zda} from the LHAPDF library~\cite{Buckley:2014ana}
with the NLO QED evolution. 
The collinear counter term is consistently implemented in the $\overline{\mbox{MS}}$ factorization scheme.

On the other hand, the incoming electron has no internal structure to accommodate for the singularity from
collinear radiations. In principle, one could keep $m_e$ throughout the calculation and retain the dependence of
the hard scattering amplitude on $\ln m_e$. This collinear singularity is physical and would break the convergence of the perturbative
calculation if $m_e$ were too small. Fortunately, the size of the fine structure constant at the electro-weak scale
ensures the convergence of the perturbation series in most of the kinematic regions we will consider (see Sec.~\ref{Pheno}).  
However, in reducing the tensor integrals
we find that some of the pentagon diagrams are sensitive to the electron mass, which leads to a significant disagreement
among different reduction programs when $m_e$ takes a small but non-zero value. This numerical instability is cured if $m_e$
is neglected and the mass singularity is regulated dimensionally.

With this observation, we factor out the mass singularity from the collinear photon radiation off the incoming electron. The singular terms are then
 put in a lepton distribution function, with the form to $\mathcal{O}(\alpha)$ in the $\overline{\mbox{MS}}$ 
 factorization scheme given by~\cite{Liu:2021jfp} (Here and below, $\alpha$ can be $\alpha_{G_{\mu}}$ or $\alpha(M_Z)$, 
 depending on the renormalization scheme used)
\begin{equation}
\begin{aligned}
f_e(x)=\delta(1-x)+\frac{\alpha}{2\pi}\Big[\frac{1+x^2}{1-x}\ln\frac{\mu_F^2}{(1-x)^2m_e^2}\Big]_+.
\label{eq:electron_pdf}
\end{aligned}
\end{equation}
This leaves a hard scattering part independent of the electron mass. Following this
 path, we are able to avoid the numerical problem during the reduction procedure
 and the calculation of the hard scattering cross section can be greatly simplified.
 
\subsubsection{4-particle final states}
\label{4-part}

\begin{figure}[H]
\centering
    \includegraphics[scale=0.5]{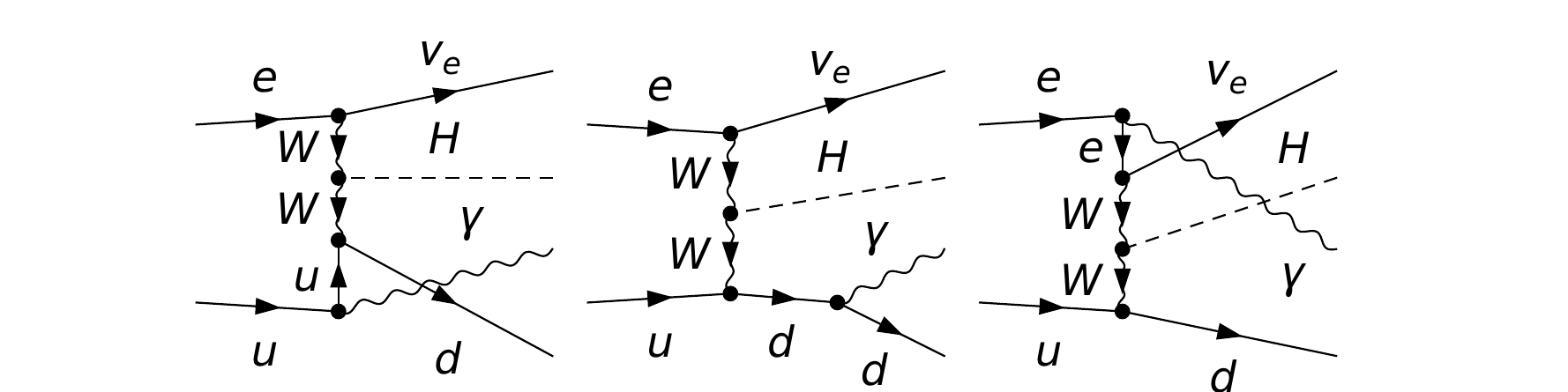} \qquad
    \includegraphics[scale=0.5]{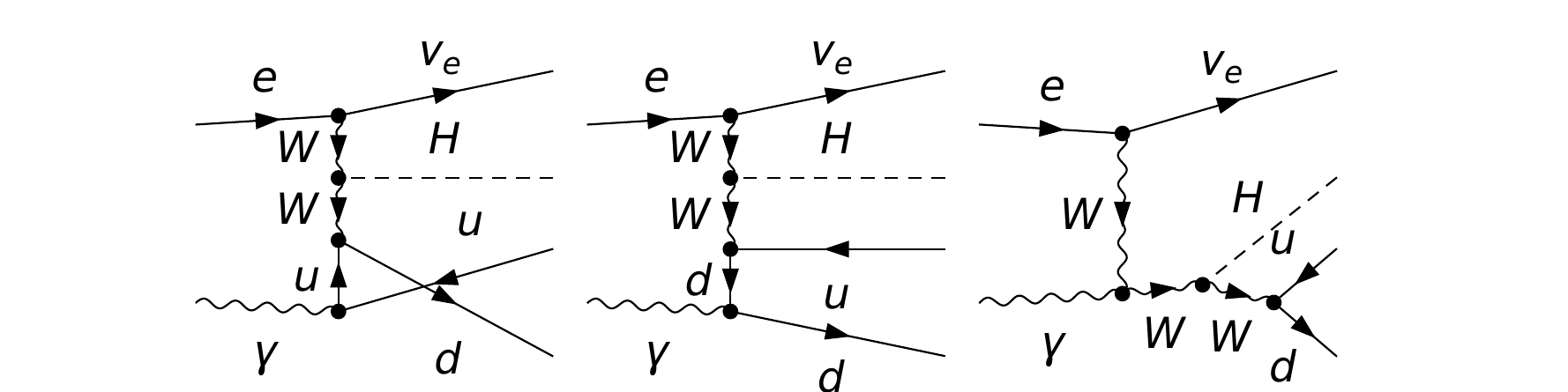}\qquad
\caption{Representative real-emission diagrams for CC WBF at the LHeC. The first and second lines 
	correspond respectively to the processes $e^-+q \rightarrow \nu_e+\gamma+q'+H$ 
	and $e^-+\gamma \rightarrow \nu_e+q+q'+H$, in which $q=u$, $q'=d$.}
\label{real}
\end{figure}

4-particle final states with an on-shell Higgs at the parton level can be produced at NLO ($O(\alpha^4)$) via  
$e^-+q \rightarrow \nu_e+\gamma+q'+H$ and $e^-+\gamma \rightarrow \nu_e+q+q'+H$. Example diagrams for these processes are given in Fig.~\ref{real}.
There can be resonant $W$ propagators, as shown by the last diagram in the figure.
To handle such cases, we employ the complex-mass scheme~\cite{Denner:1999gp,Denner:2005fg,Denner:2006ic} in the calculation of the amplitudes.
The subtraction procedure starts by constructing terms that asymptotically approach the squared amplitude for the real 
emission process in the phase space region where the radiated particle (the emittee) becomes soft or 
collinear to the emitter. The dipole structure of the subtraction term is encoded in the charge correlation 
between the splitting pair and a third particle (spectator).
Three types of dipoles are present for our processes and the subtraction term takes the form~\cite{Schonherr:2017qcj}
\begin{equation}
\begin{aligned}
|\mathcal{M}_{sub}|^2=\sum_{i, j}\sum_a\mathcal{D}_{ij}^a+\sum_{a,j}\sum_{k\neq j}\mathcal{D}_{j,k}^a+\sum_{a,j}\sum_{b\neq a}\mathcal{D}_{j}^{a,b}.
\label{eq:realdip}
\end{aligned}
\end{equation}
The indices $i$, $j$, $k$ denote the emitter, emittee, and spectator, respectively, in the final state.  
$a$ and $b$ are initial state partons, where $a$ plays the role of 
the spectator in $\mathcal{D}_{ij}^a$, and the emitter in $\mathcal{D}_{j,k}^a$ and $\mathcal{D}_{j}^{a,b}$, while $b$ is the spectator.
Here we extend the meaning of ``parton'' to include the initial state electron and photon that undergo electromagnetic interactions.
The outer sum is over all possible choices of the splitting pairs, while the inner sum runs over the spectators. The diagrammatic representation
of the three dipole types can be found in Ref.~\cite{Catani:1996vz,Schonherr:2017qcj}.
In each dipole term, the singular behavior is extracted and encoded in a splitting function  factorized from 
the corresponding born process, whose final state coincides with the kinematics of the radiative process in the soft or collinear limit.
To obtain the born kinematics one combines the emitter and emittee from the corresponding splitting process
 to a single particle. The momenta of this particle and of the spectator are then shifted in such a way as to 
 respect momentum conservation and on-shell conditions for the 3-particle final state~\cite{Catani:1996vz}.
The explicit expressions for the dipole terms and shifted momenta are given in Appendix~\ref{app:4-particle-final-states}

Now we subtract these terms from the real emission part to remove any singularity 
in the phase space. The cross section is thereby expressed by
\begin{equation}
\begin{aligned}
\sigma_4^{NLO}=\sum_b \int_0^1d\eta_bf_b(\eta_b,\mu_F)\frac{1}{2\hat{s}}d\Phi_4\left\{|\mathcal{M}_R^b|^2F^{(2)}(p_1,p_2,p_3,p_4;P_A,p_b)-|\mathcal{M}_{sub}^b|^2\right\},
\label{eq:realsub}
\end{aligned}
\end{equation}
where $b$ denotes the parton that collides with the electron; it can be a quark, anti-quark, or photon, 
and plays the role of either an emitter or spectator. 
It should not be confused with the ``$b$'' in Eq.~\ref{eq:realdip}, which only denotes a spectator.

In order to construct collinear-safe observables, we assign a jet function $F^{(2)}$ to the real emission process
to combine quarks and photons according to a certain choice of the jet algorithm. $F^{(2)}$ depends on the momenta of the
4-particle final state. Correspondingly, for each dipole term in Eq.~\ref{eq:realdip}, there is a function $F^{(1)}$ of
the 3-particle phase space obtained from the real emission process via the procedure described after Eq.~\ref{eq:realdip}. 
$F^{(1)}$'s are implicitly included in Eqs.~\ref{eq:realdip} and ~\ref{eq:realsub}, as well as in Eq.~\ref{eq:LOxsec}.
To ensure the cancellation of singularities, $F^{(2)}$ must approach the corresponding $F^{(1)}$ in various soft and collinear limits.

 At this order, only the delta function term of the electron PDF in Eq.~\ref{eq:electron_pdf} contributes, 
which is trivially integrated out in Eq.~\ref{eq:realsub} over the electron momentum fraction (This is also true for the LO case in Eq.~\ref{eq:LOxsec}).
Since the integrand in Eq.~\ref{eq:realsub} is finite, integration over the 4-particle phase space
can be safely done in four dimensions with numerical methods.

\subsubsection{3-particle final states}

Evidently, the subtracted dipole terms are auxiliary and need to be put back to avoid any artifact in the calculation.
The structure of these terms is simple enough such that the singularities 
can be isolated in a factorized form after an analytical integration over a one-particle phase space. 
These ``integrated'' dipole terms have a phase space with one fewer particle, and are added back to combine with the loop diagrams.
The result contains only collinear singularities from the initial state radiations, which are then canceled by the collinear counter terms
from the factorization procedure.

The dipole and collinear counter terms can be reorganized into a term $\boldsymbol{I}$ that contains all the soft and 
collinear divergences present in the loop diagrams, and two other terms  $\boldsymbol{K}$ and $\boldsymbol{P}$ that have finite remainders after 
cancellation of singularities. In particular, $\boldsymbol{K}$ and $\boldsymbol{P}$ include the finite terms from factorization 
of initial state collinear singularities. The terms that depend on the choice of factorization scheme 
are contained in $\boldsymbol{K}$, while $\boldsymbol{P}$ gives the dependence on the factorization scale $\mu_F$.
The cross section from the 3-particle final states can therefore be 
expressed in terms of the contribution from the born and loop processes together with 
$\boldsymbol{I}$,$\boldsymbol{K}$ and $\boldsymbol{P}$~\cite{Catani:1996vz,Catani:2002hc,Schonherr:2017qcj} 
\begin{equation}
\begin{aligned}
	\sigma_3^{NLO}=&\sum_{b}\int  d\eta_b f_b(\eta_b,\mu_F)\bigg\{\int \frac{1}{2\hat{s}}d\Phi_3^{(4)}\Big[V_{ab}(\Phi_3,P_A,p_b)+B_{ab}(\Phi_3,P_A,p_b) \boldsymbol{I}^b(\epsilon,\mu^2)\Big]_{\epsilon =0}\\
&+\sum_{a'}\int dx_a\int\frac{1}{2\hat{s}}d\Phi_3^{(4)} B_{a'b}(\Phi_3^{(4)},x_a P_A,p_b) \Big[\boldsymbol{K}^b_{aa'}(x_a)+\boldsymbol{P}^b_{aa'}(x_a;\mu_F^2)\Big]\\
	&+\sum_{b'}\int dx_b\int\frac{1}{2\hat{s}}d\Phi_3^{(4)} B_{ab'}(\Phi_3^{(4)},P_A,x_b p_b) \Big[\boldsymbol{K}^a_{bb'}(x_b)+\boldsymbol{P}^a_{bb'}(x_b;\mu_F^2)\Big]\bigg\}\\
	&+\sum_{b}\int d\eta_a d\eta_b f^{\mathcal{O}(\alpha)}_e(\eta_a,\mu_F)f_b(\eta_b,\mu_F)\int \frac{1}{2\hat{s}}d\Phi_3^{(4)}B_{ab}(\Phi_3^{(4)},p_a,p_b).
\label{eq:virtualsub}
\end{aligned}
\end{equation}
Here $B$ is the squared born amplitude, and $V$ the interference between born and loop processes. 
$a$ labels the electron and $b$ the parton from the proton; 
$a'$ ($b'$) is the flavor from the splitting of $a$ ($b$) that enters the LO hard scattering amplitude. 
The beam and initial state parton momenta are related by $p_a=\eta_a P_A$ and $p_b=\eta_b P_B$.
In the last line of Eq.~\ref{eq:virtualsub}, we use the $\mathcal{O}(\alpha)$ term of the electron PDF, as is indicated by its superscript.
The first three lines are from convolution with the delta term in Eq.~\ref{eq:electron_pdf}, for which $\eta_a = 1$.
In each term on the R.H.S of Eq.~\ref{eq:virtualsub}, the phase space $\Phi_3^{(4)}$ and flux factor $1/2\hat{s}$ 
both depend on the CM energy of the corresponding born factor.
The subscripts of $B$ and $V$ in the first and last lines show that the flavors entering 
these factors are the same as those from the incoming beams (i.e., from the PDFs). In contrast, the subscripts of $\boldsymbol{K}$ and $\boldsymbol{P}$
show that the flavors may change before and after initial state splittings in these terms. The superscript of $\boldsymbol{K}$ or $\boldsymbol{P}$
labels the other incoming particle that does not go through the splitting.
  As in Eqs.~\ref{eq:LOxsec}, ~\ref{eq:realdip} and ~\ref{eq:realsub}, 
 we implicitly include a jet function $F^{(1)}$ for the phase space of each term in Eq.~\ref{eq:virtualsub}.

The loop integrals in $V$
are done in $d = 4- 2\epsilon$ dimensions and all singularities after renormalization manifest as single and double poles, to be
canceled exactly by the poles from $\boldsymbol{I}$. Therefore, the phase space $\Phi_3$ of $V$ and $B$ in the first line of Eq.~\ref{eq:virtualsub} 
should be the same, i.e., both in either 4 or $d$ dimensions. After the poles are canceled, one takes the limit $\epsilon \rightarrow 0$ and 
performs the phase space integration in 4 dimensions, as denoted by ``$\epsilon=0$'' and ``$d\Phi_3^{(4)}$'' in the first line.
While $\boldsymbol{I}$,$\boldsymbol{K}$ and $\boldsymbol{P}$ all contribute to the quark/anti-quark induced processes,
 photon induced processes only receive contribution from non-singular terms $\boldsymbol{K}$ and $\boldsymbol{P}$.

The expressions of $\boldsymbol{I}$,$\boldsymbol{K}$ and $\boldsymbol{P}$ for specific processes  
are listed in Appendix~\ref{app:3-particle-final-states}. The integrand of all terms in 
Eq.~\ref{eq:virtualsub} are finite and the phase space can be integrated over numerically in four dimensions.
Once  $\sigma_3^{NLO}$ and $\sigma_4^{NLO}$ are obtained, we combine them to give the full corrections to the cross section 
at NLO. Note that neither of the two contributions alone is physical because of the auxiliary dipole terms introduced.

\section{Numerical result}
{\label{pheno}}
\subsection{Setup}
To make numerical predictions for the WBF cross section, we take the energies of the incoming electron and proton to be
\begin{equation}
\begin{aligned}
E_e=140 \; \mbox{GeV}, \qquad E_p=7 \; \mbox{TeV},
\end{aligned}
\end{equation}
corresponding to a CM energy $\sqrt{s}=2\sqrt{E_pE_e}\approx 1.98\;\mbox{TeV}$, 
which is the choice in the study of $H$-$b$ coupling at the LHeC~\cite{Han:2009pe}.
The renormalization and factorization scales are set to $M_W$. 
In addition, we use the following set of parameters
\begin{equation}
\begin{aligned}
	&G_{\mu}=1.16639\times 10^{-5} \; \mbox{GeV}^{-2}, \qquad \alpha_{G_{\mu}}=1/132.5,\qquad \alpha(M_Z)=1/128.93,\\
	&M_W=80.419 \; \mbox{GeV}, \qquad \Gamma_W=2.09291 \; \mbox{GeV},\qquad M_Z=91.188 \; \mbox{GeV}, \\
	& \Gamma_Z=2.49877 \; \mbox{GeV},\qquad c_W^2=1-s_W^2=\frac{M_W^2}{M_Z^2},\\
    &m_e=0.510998928 \; \mbox{MeV}, \qquad M_H=125 \; \mbox{GeV}.
\end{aligned}
\end{equation}
 As discussed in Sec.~\ref{Ren} and ~\ref{Fac}, the dominant fermion mass effect is incorporated into the renormalized 
fine structure constant as well as the electron distribution. Hence we do not specify the values of fermion masses here, 
 and set them to zero throughout the calculation of the hard scattering cross section. The only exception is the explicit use of $m_e$ 
 in the electron distribution from Eq.~\ref{eq:electron_pdf}. As stated before, we use a unit CKM matrix in this work.

In this calculation, the tree level Feynman diagrams and amplitudes are produced and 
evaluated by the {\sc FeynArts}, {\sc FormCalc}, and {\sc LoopTools} package set~\cite{Hahn:2000kx,Hahn:1998yk}.
The analytical calculation of the dipole terms are carried out with the program {\it Mathematica}. The one-loop amplitudes are computed using 
the program {\sc MadLoop}~\cite{hirschi:2011pa} implemented by {\sc MadGraph5\_aMC@NLO} package~\cite{Alwall:2014hca}.
We have developed our own code to generate the 3- and 4-particle phase space, over which
the numerical integration is performed with the Vegas Monte-Carlo program implemented by the {\sc Cuba} library~\cite{Hahn:2004fe}.

\subsection{Consistency checks}
To verify the calculation we have done several checks at different levels. 
First, the tree-level amplitudes for born and real emission processes computed with {\sc FormCalc} 
are compared with those obtained from {\sc MadGraph5}. 

In the 4-particle final state part, we have made sure that the real emission 
contribution approach the corresponding dipole terms in various soft and collinear limits, 
and that the integrand after subtraction be stable in a Monte-Carlo integration. Also, the explicit form of the dipole terms
for the real emission processes is compared numerically with the dipoles generated by {\sc MadDipole}~\cite{Frederix:2008hu,Gehrmann:2010ry}.

For the 3-particle final state terms, we have verified analytically in the case of triangle graphs that all double and 
single poles produced in dimensional regularization are canceled among the loop graphs 
(whose forms are derived from the scalar integrals in Ref.~\cite{Ellis:2007qk} after reduction), dipoles, and the collinear counter terms. 
This verification is then carried out for all loop diagrams (computed with {\sc MadLoop}) numerically. In the finite part, we checked that the dependence on the 
renormalization scale is canceled between the loop and dipole terms.\footnote{Note that the factorization scale independence could not be 
checked in this calculation because the PDF sets undergo QCD evolution. The corresponding QCD corrections need to be included in order to 
cancel the scale dependence at the order of evolution.} There is no running of the coupling and masses in the renormalization 
schemes we choose, and the cancellation of the scale dependence is exact at the order we are working with.

\subsection{Phenomenology}
\label{Pheno}

First, we report the result of the integrated cross section computed in two renormalization schemes. 
Table.~\ref{int-xsec} lists contribution from LO and NLO terms, where all $W$ fusion processes that produce a Higgs boson are included.
\begin{table}[h]
\begin{center}
\begin{tabular}{|c|c|c|c|c|}
\hline
	Schemes & total & LO & 3-particle & 4-particle \\
\hline
	$G_{\mu}$ &  223.27 & 245.48 &  -20.48 & -1.73 \\
\hline
	$\alpha_{M_Z}$ & 216.74 & 266.48 & -47.82 & -1.93 \\
\hline
\end{tabular}
\end{center}
	\caption{Integrated cross sections in fb for CC WBF at the LHeC at LO and NLO, computed in two renormalization schemes $G_{\mu}$ and $\alpha_{M_Z}$. The electron and proton beam energies are $140$~GeV and $7$~TeV, respectively.}
\label{int-xsec}
\end{table}
There is a prominent difference in the LO cross sections obtained in the two schemes. This is solely due to the difference in the renormalized
fine structure constants. In the $G_{\mu}$ scheme, the NLO terms reduces the LO result by $\sim 9\%$, while the relative correction in the
$\alpha(M_Z)$ scheme is as large as $\sim 18\%$. The sum of LO and NLO result, however, agree in both schemes, up to a difference at one higher order.
This in fact provides another consistency check for  the calculation. In both schemes the bulk of the NLO corrections come from the 
3-particle final states. The allocation of the contribution between 3- and 4- particle final states is  an artifact that relies on
the choice of the subtraction terms away from the divergent region of the phase space. But this will of course not affect the sum of the
two contributions.
Notice that the coupling in the $G_{\mu}$ scheme is smaller and leads to a much faster convergence of the perturbation series. In the following
discussion we shall stick with this scheme. 

Next, we turn to the study of distributions. 
As remarked in the introduction, a distinctive feature of the WBF event in $e$-$p$ collisions is an energetic 
jet produced in the far backward direction with large transverse momentum (we take the direction of the incoming electron to be forward). 
This is very prominent at the LHeC because the final state is strongly boosted along the direction of the proton. 
The decay product of the Higgs tends to be well separated from this jet, with a rapidity near the central region. 
 To make full use of this event shape, we follow the strategy in Ref.~\cite{Han:2009pe,Jager:2010zm} 
 and identify the ``tagging jet'' as the one with the largest transverse momentum, imposing the cuts
\begin{equation}
\begin{aligned}
\label{eq:cuts}
	p_{T}^{j}>30~\text{GeV}, \qquad  -5<\eta_{j}<-1 , \qquad M_{H,j}>250~\text{GeV},
\end{aligned}
\end{equation}
where $p^j_T$ and $\eta^j$ are the transverse momenta and rapidity of the tagging jet; 
$M_{H,j}$ is the invariant mass of the Higgs-tagging jet system. 
Further, the large momentum transfer to the neutrino from the hard scattering is reflected by the requirement
\begin{equation}
\begin{aligned}
\label{eq:cuts}
\slashed{E}_{T}>25~\text{GeV},
\end{aligned}
\end{equation}
with $\slashed{E}_{T}$ denoting the transverse energy of the neutrino.
Normally one also applies basic cuts on the Higgs decay products,
which is apparently not done in our case as we are not performing a full signal-background analysis in this work.

In the following, we present distributions in various observables $\mathcal{O}$ computed at LO 
and NLO (here ``NLO'' also includes the LO contribution). 
Each jet function in Eqs.~\ref{eq:realsub} and ~\ref{eq:virtualsub} plays the role of 
constructing the observable with the momenta from its argument list.\footnote{Note that the momenta for a dipole term are transformed from those of
the corresponding real emission process.} Here we use $k_T$ algorithm~\cite{Catani:1992zp,Ellis:1993tq,Blazey:2000qt} 
with the parameter $D=0.8$ to define jet observables.
In order to show the relative correction from the EW radiations with respect to the LO result, we define the $K$ factor
\begin{equation}
\begin{aligned}
	K(\mathcal{O})\equiv \frac{d\sigma^{NLO}/d\mathcal{O}}{d\sigma^{LO}/d\mathcal{O}},
\end{aligned}
\end{equation}
as a function of the observable $\mathcal{O}$.

\begin{figure}[h]
\centering
\subfloat[$p_T^{j}$ distribution]{\includegraphics[scale=0.3]{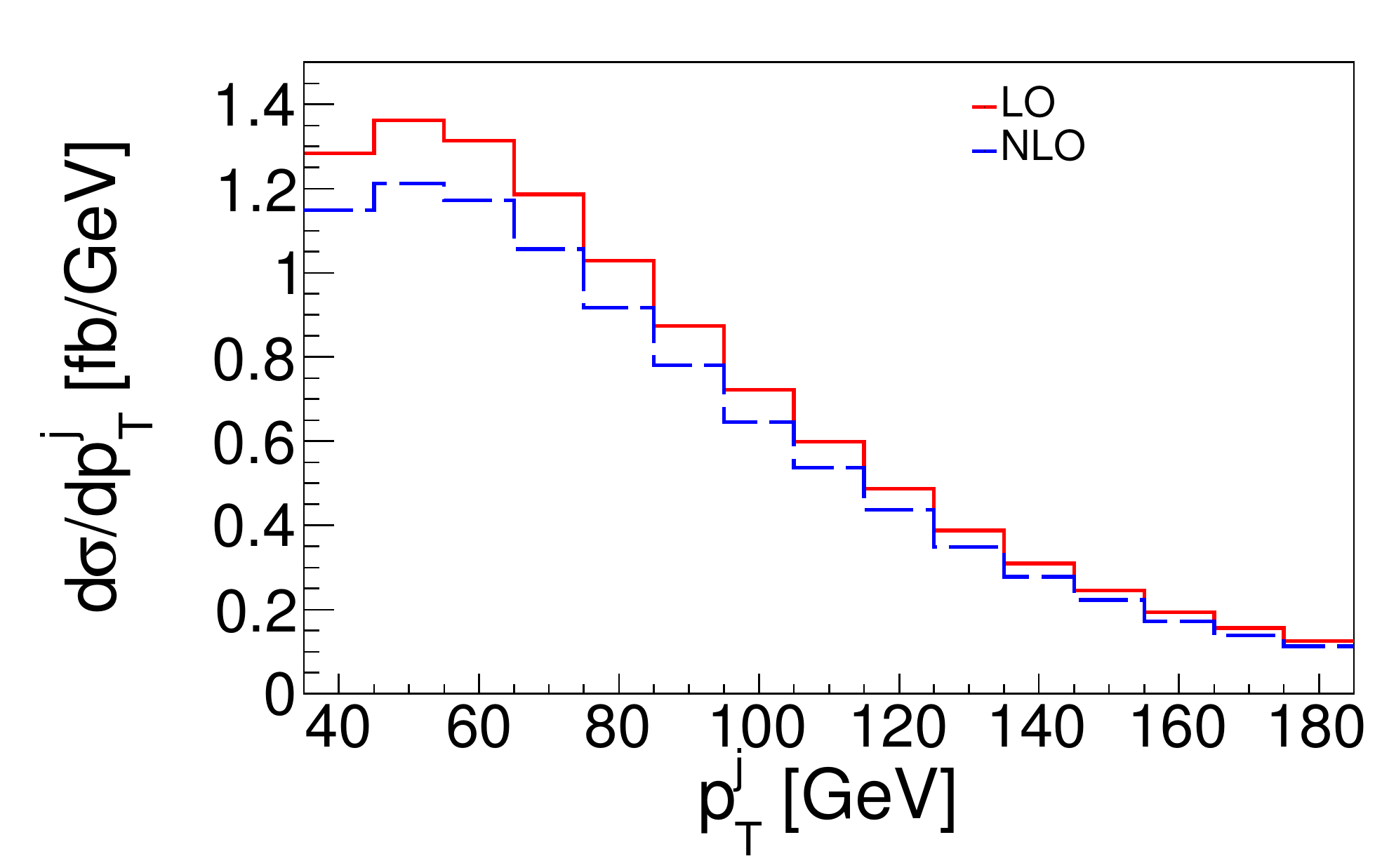}}\qquad
\subfloat[$K$ factor]{\includegraphics[scale=0.3]{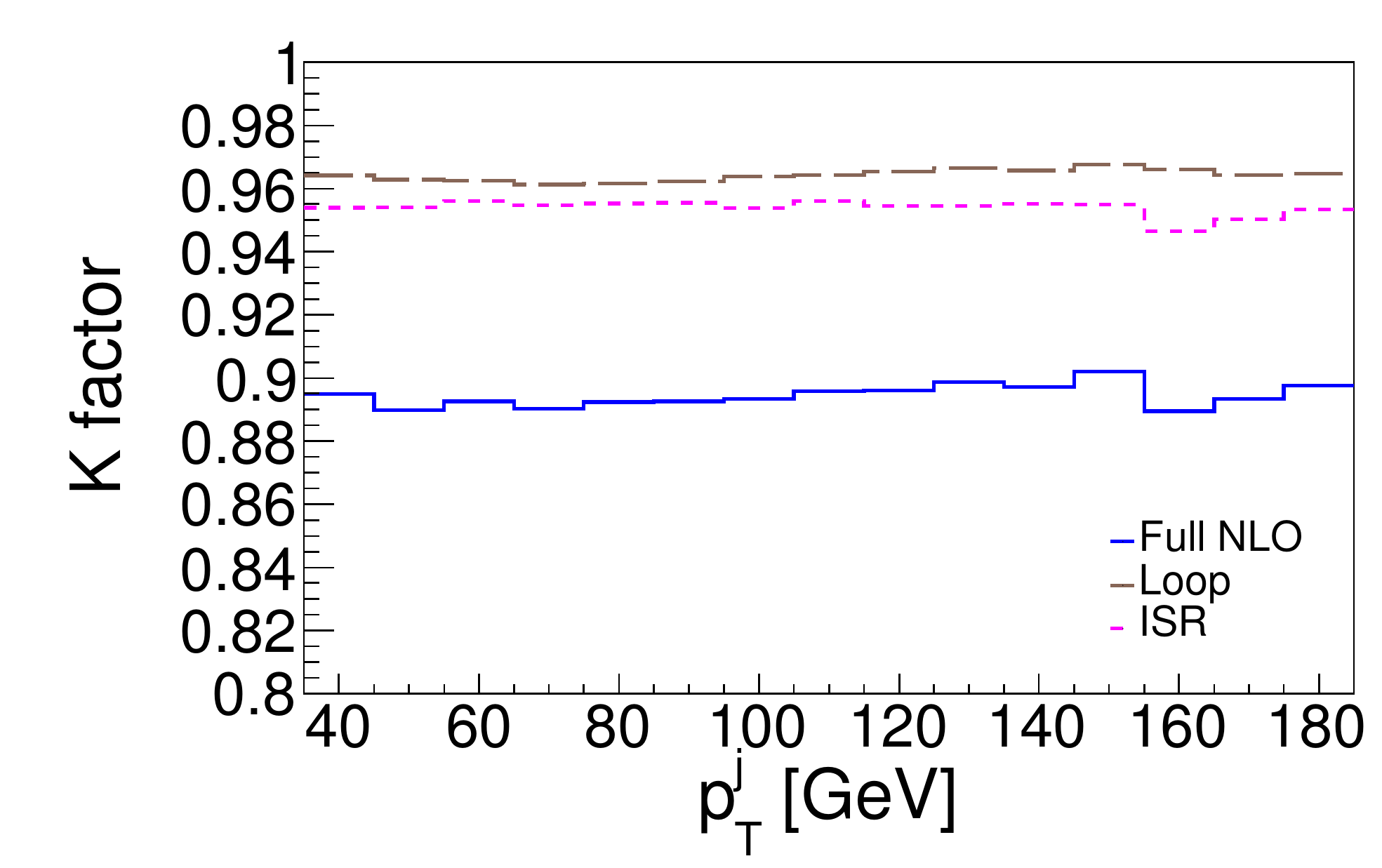}}\qquad \\
	\caption{Distribution in the transverse momentum $p_T^j$ of the tagging jet (a), and the corresponding $K$ factors (b).}
\label{ptj}
\end{figure}
 We first display the differential distribution  of the tagging jet transverse momentum $p_T^j$ in Fig.~\ref{ptj} (a),  where the LO and NLO curves
 are in red and blue respectively. 
The cross section peaks near small value of $p_T^j$ and drops apace with increasing jet transverse momentum, with an average $p_T^j$ of order
the $W$ boson mass. 
The relative EW correction is about -10\% over the broad $p_T^j$ range, as shown by the blue curve on the right. In addition, we show separately
the $K$ factor of the two dominant contributions at NLO. The short-dashed magenta line gives the correction from the 
initial state radiation (ISR) of the photon off the electron, which is about -5\% and is consistent with the result 
obtained by Ref.~\cite{Blumlein:1992eh}. This collinear radiation results in
large logarithmic terms of the form $\sim \ln m_e$ (See the last line of Eq.~\ref{eq:virtualsub}), and gives the main part of the QED
corrections. Another contribution in long-dashed brown is from the EW loop diagrams, with IR singular terms removed by subtractions.
In Fig.~\ref{ptj} it is of the similar size as ISR, and both curves have a rather flat shape.

\begin{figure}[h]
\centering
\subfloat[$\eta_{j}$ distribution]{\includegraphics[scale=0.3]{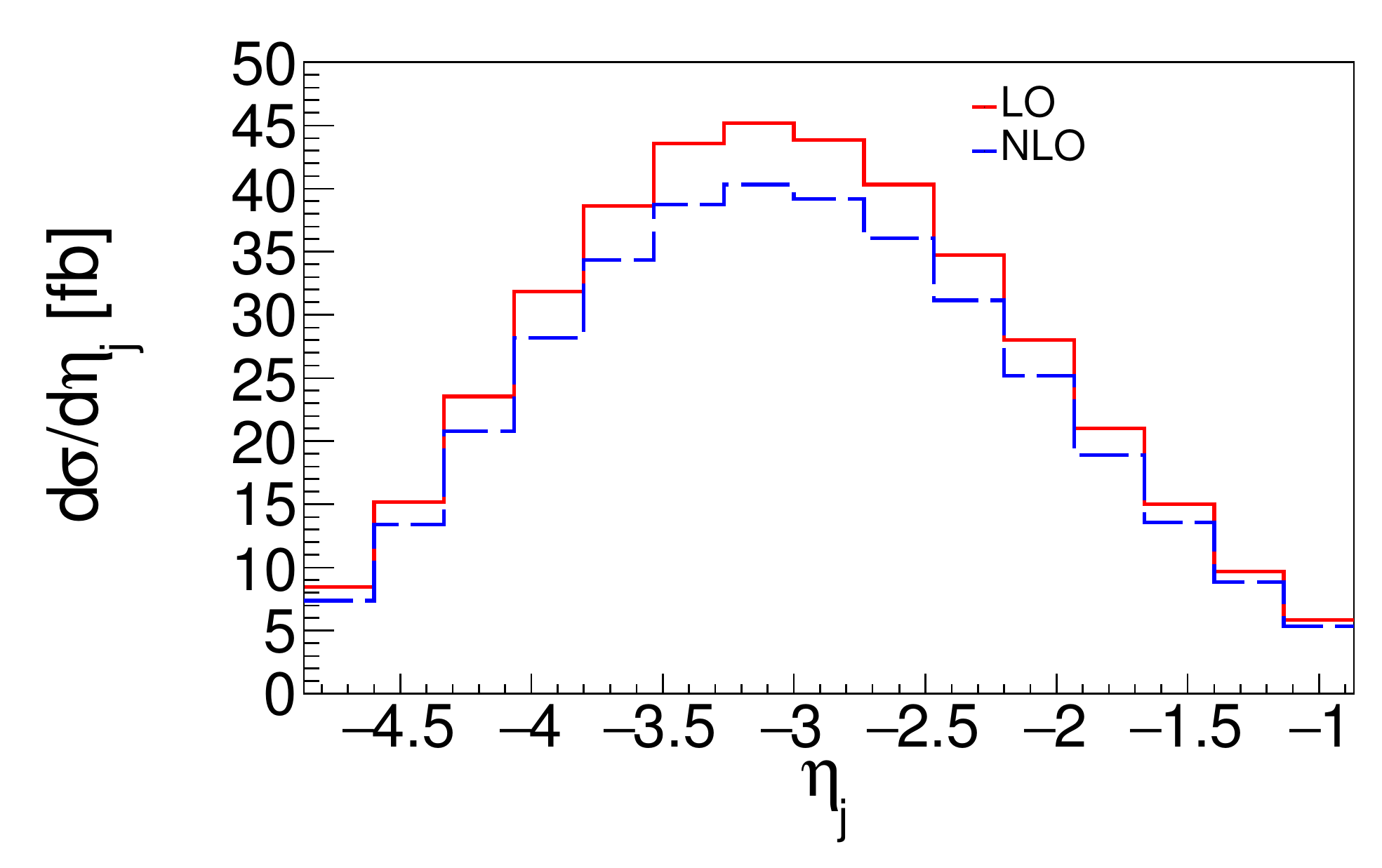}}\qquad
\subfloat[$K$ factor]{\includegraphics[scale=0.3]{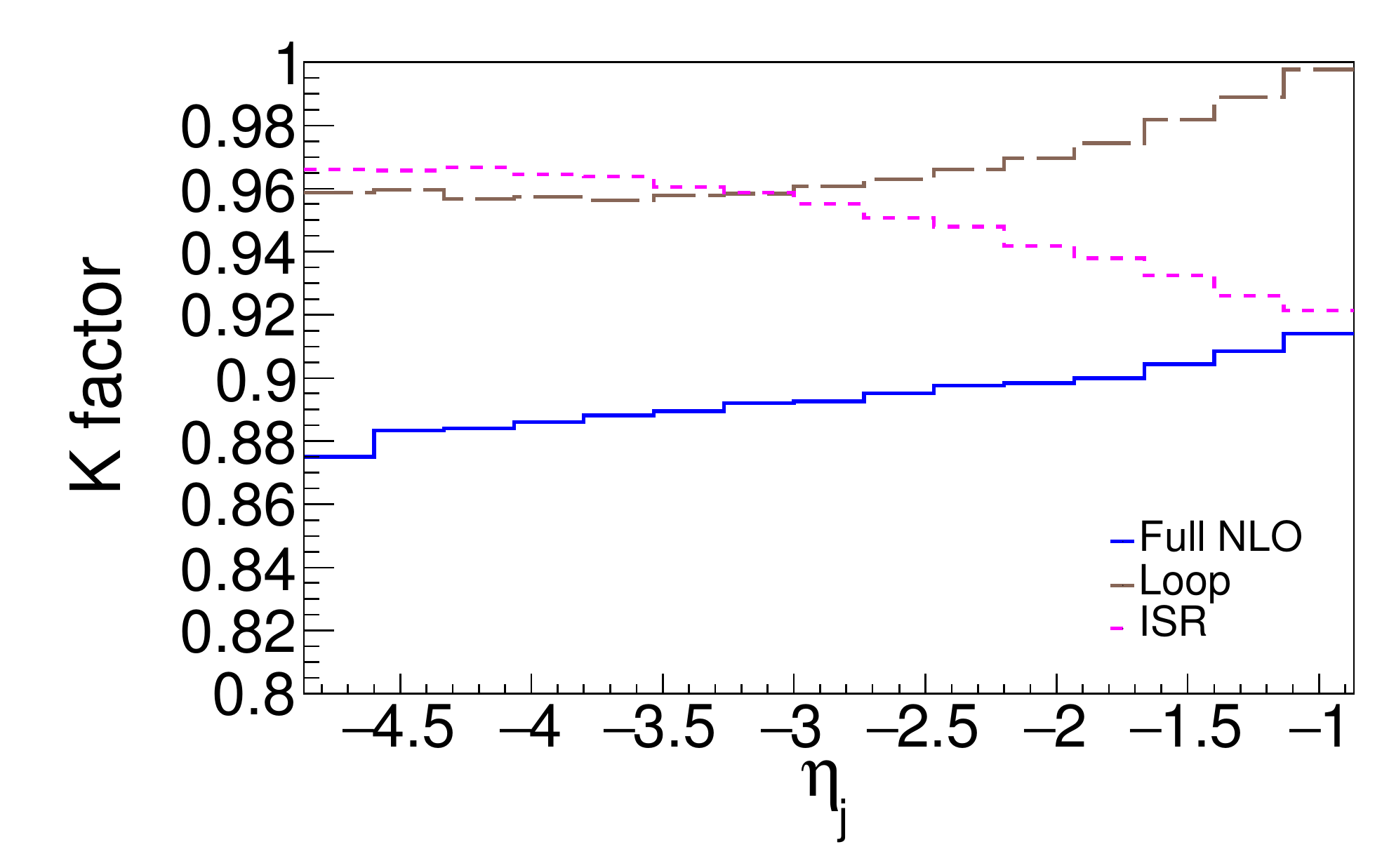}}\qquad \\
	\caption{Distribution in the rapidity $\eta_j$ of the tagging jet (a), and the corresponding $K$ factors (b).}
\label{etaj}
\end{figure}
The distribution in the rapidity $\eta_j$ of the tagging parton is shown in Fig.~\ref{etaj}, 
where the peak is clearly located in the backward region. The EW corrections vary between -8\% and -12\%, with a slow increase as $\eta_j$
approaches the central region. The loop contribution increases more rapidly at large $\eta_j$ than the full NLO curve, while the ISR terms 
display an opposite trend in this region.

\begin{figure}[h]
\centering
\subfloat[$p_T^{H}$ distribution]{\includegraphics[scale=0.3]{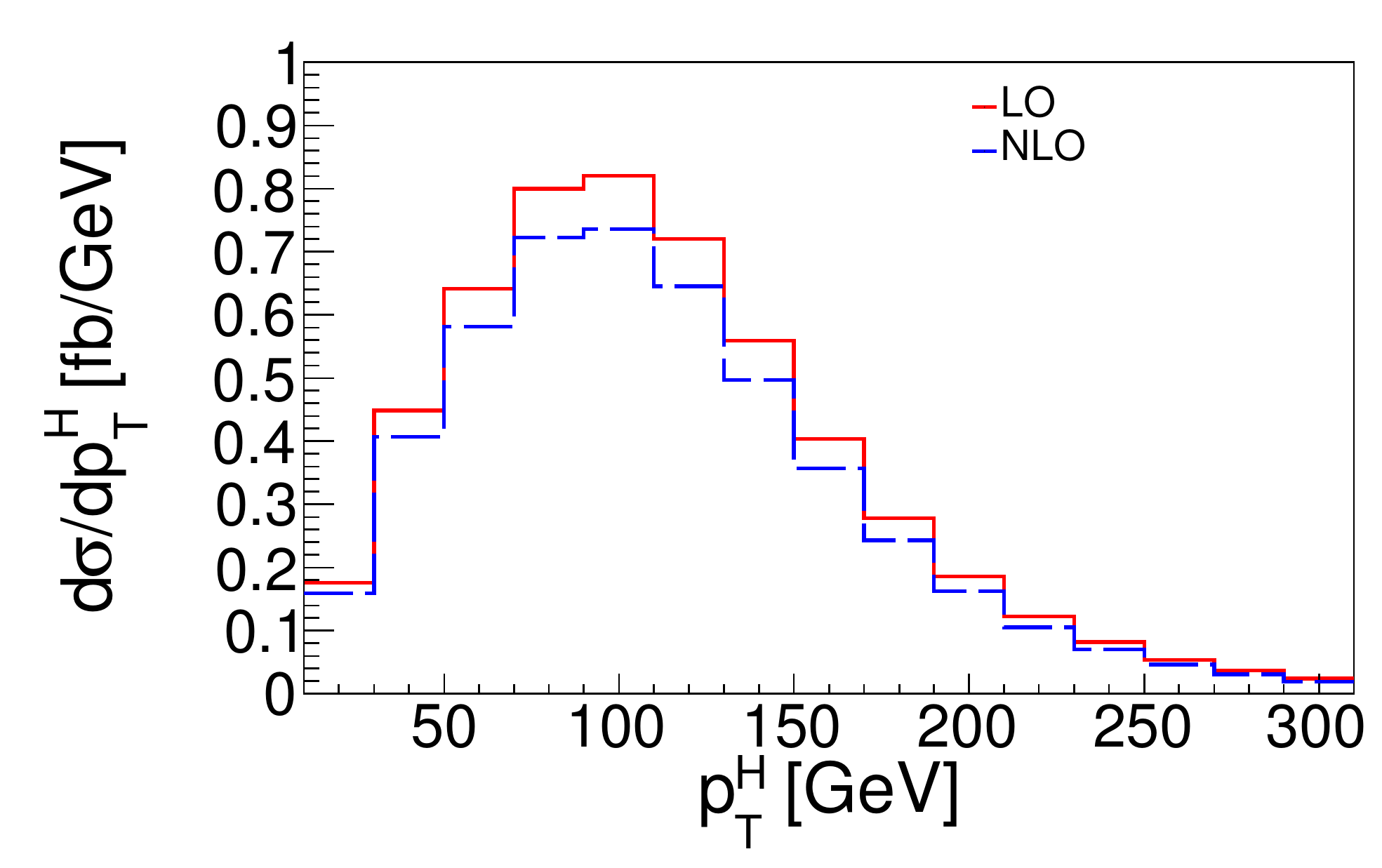}}\qquad
\subfloat[$K$ factor]{\includegraphics[scale=0.3]{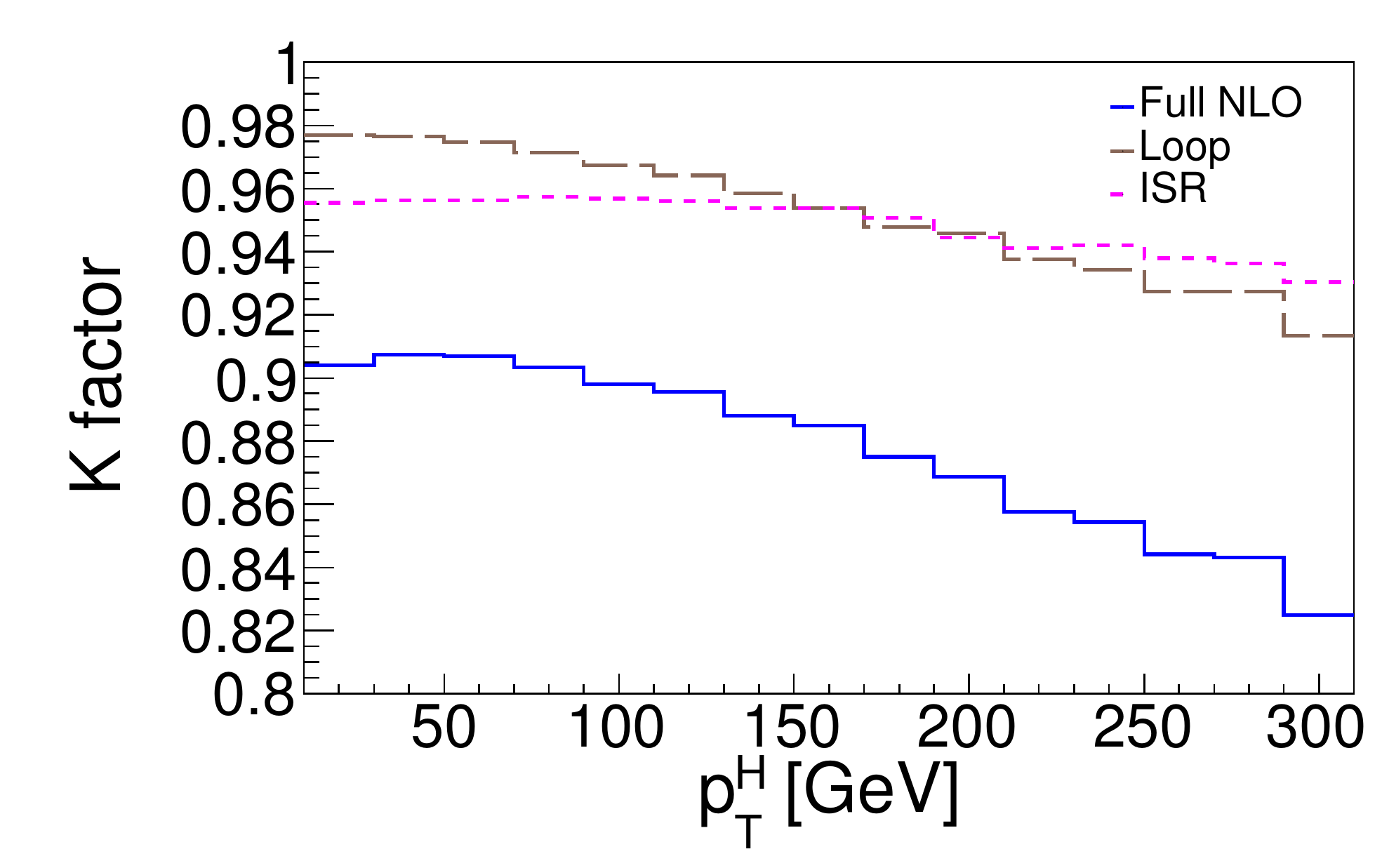}}\qquad \\
	\caption{Distribution in the transverse momentum $p_T^H$ of the Higgs boson (a), and the corresponding $K$ factors (b).}
\label{pth}
\end{figure}
\begin{figure}[h]
\centering
\subfloat[$\eta_{H}$ distribution]{\includegraphics[scale=0.3]{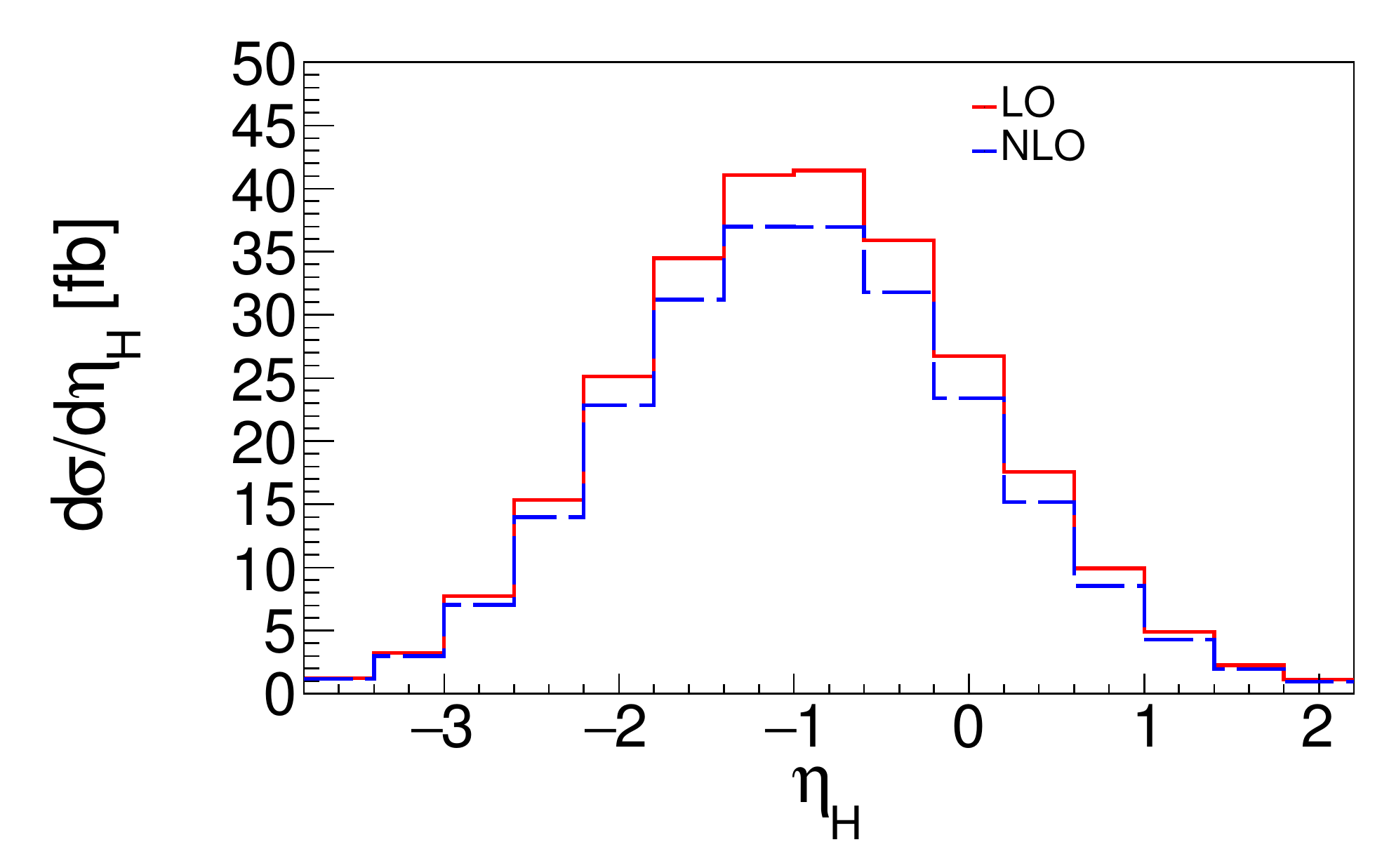}}\qquad
\subfloat[$K$ factor]{\includegraphics[scale=0.3]{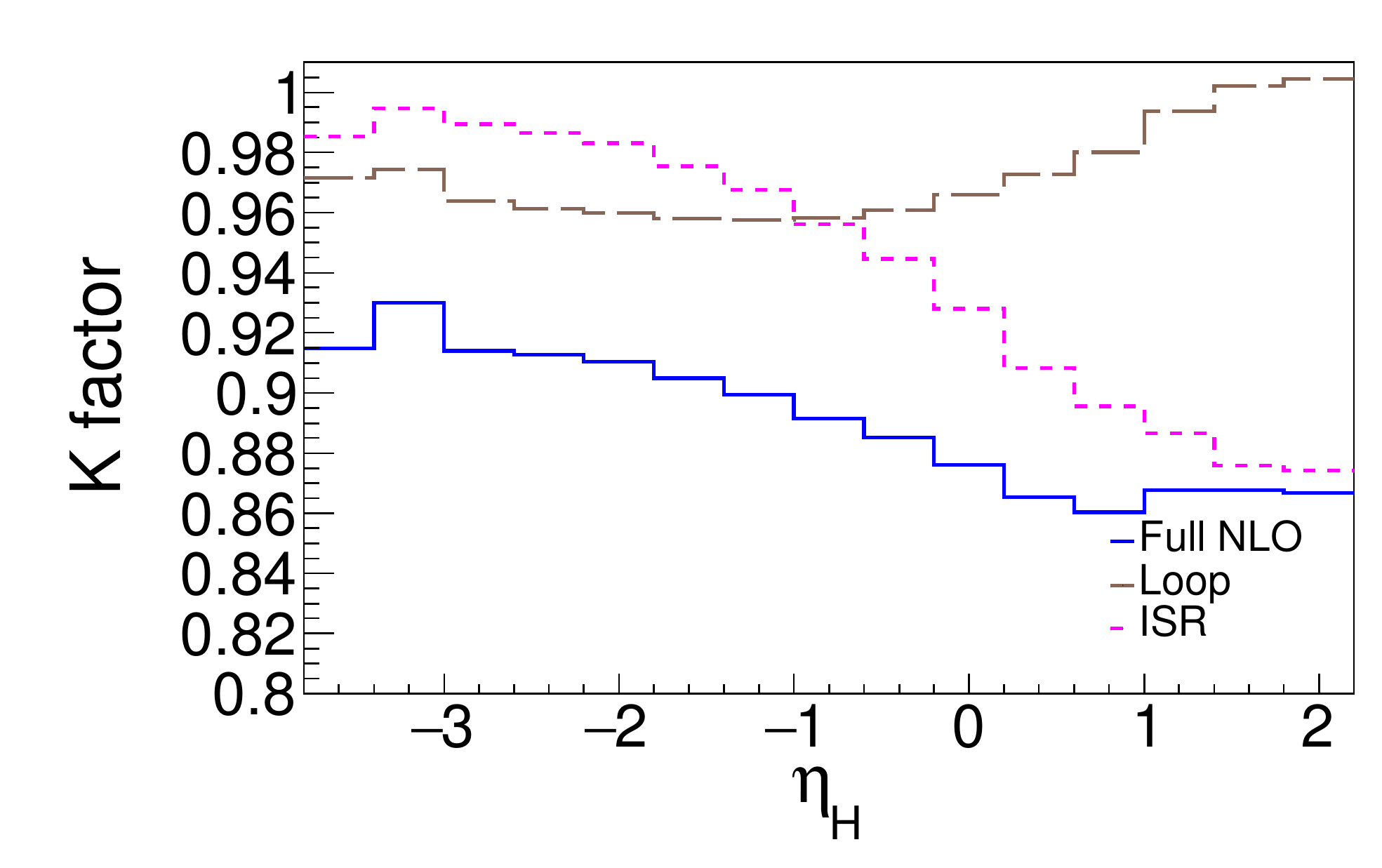}}\qquad \\
	\caption{Distribution in the rapidity $\eta_H$ of the Higgs boson (a), and the corresponding $K$ factors (b).}
\label{etah}
\end{figure}
Figs.~\ref{pth} and ~\ref{etah} show that the Higgs observables (transverse momentum $p_T^H$ and rapidity $\eta_H$) are more sensitive to the
NLO terms. The size of the corrections can be as large as  -17\% (for $p_T^H$) or -14\% (for $\eta_H$). There are also considerable changes
of the $K$ factors in the ranges of both variables. In particular, ISR terms play a prominant role in the $\eta_H$ distribution. It is 
argued~\cite{Denner:2003ri,Denner:2004jy} that the ISR contribution can be enhanced near the threshold for producing the final state. 
This is already seen at large $p_T^H$ or positive $\eta_H$ in the plots. 
The analysis in Ref.~\cite{Han:2009pe} shows that the WBF process tends to produce the final state
with a larger invariant mass than that of the background. This may induce a heavier suppression of the WBF signal from the ISR terms 
in the region where the signal dominates,  potentially making the probe of e.g. $H$-$b$ Yukawa more difficult.
Of course, radiative corrections for the background need also be accounted for in order to yield a full phenomenological effect.

\begin{figure}[h]
\centering
\subfloat[$\Delta\phi_{MET-j}$ distribution]{\includegraphics[scale=0.3]{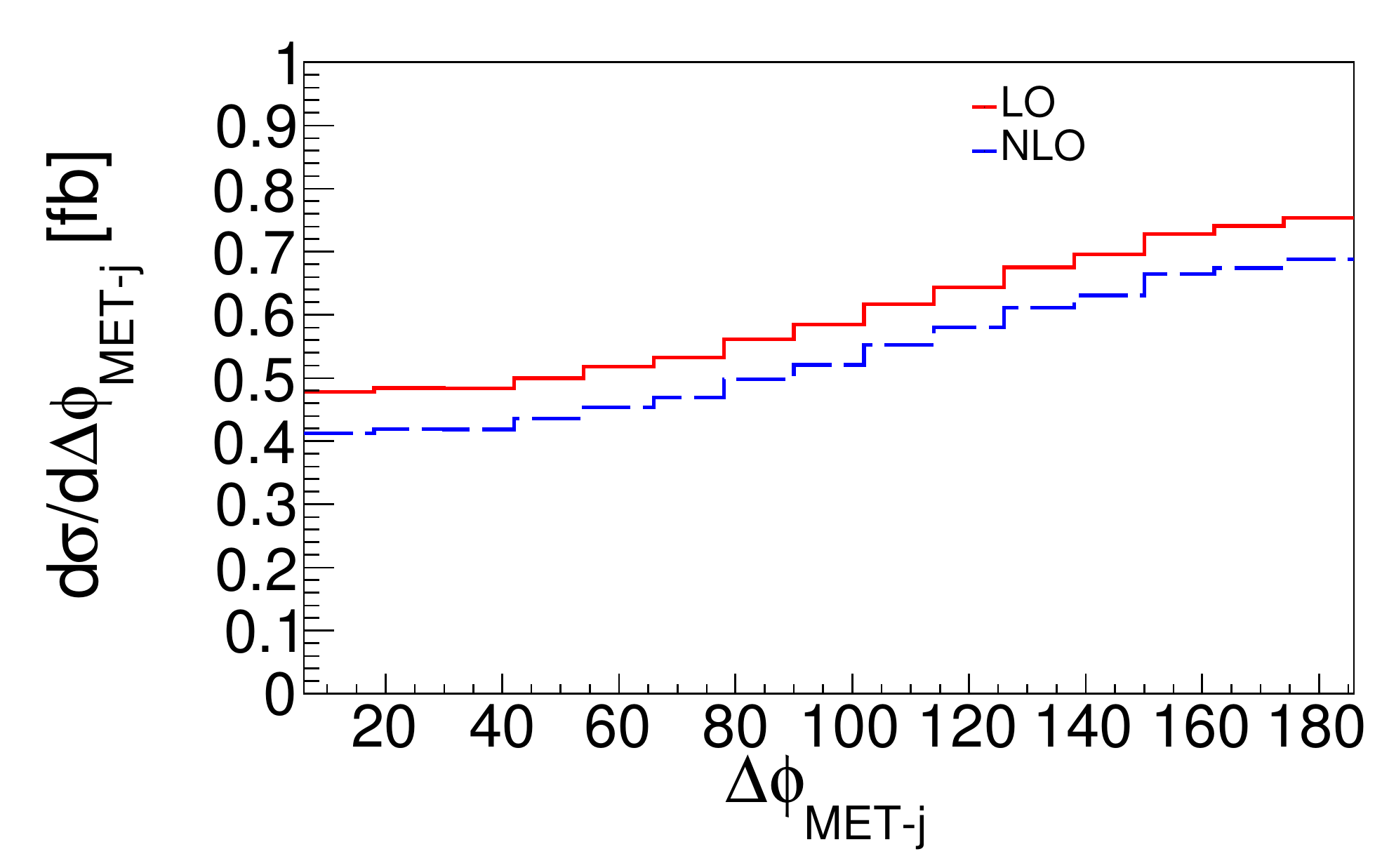}}\qquad
\subfloat[$K$ factor]{\includegraphics[scale=0.3]{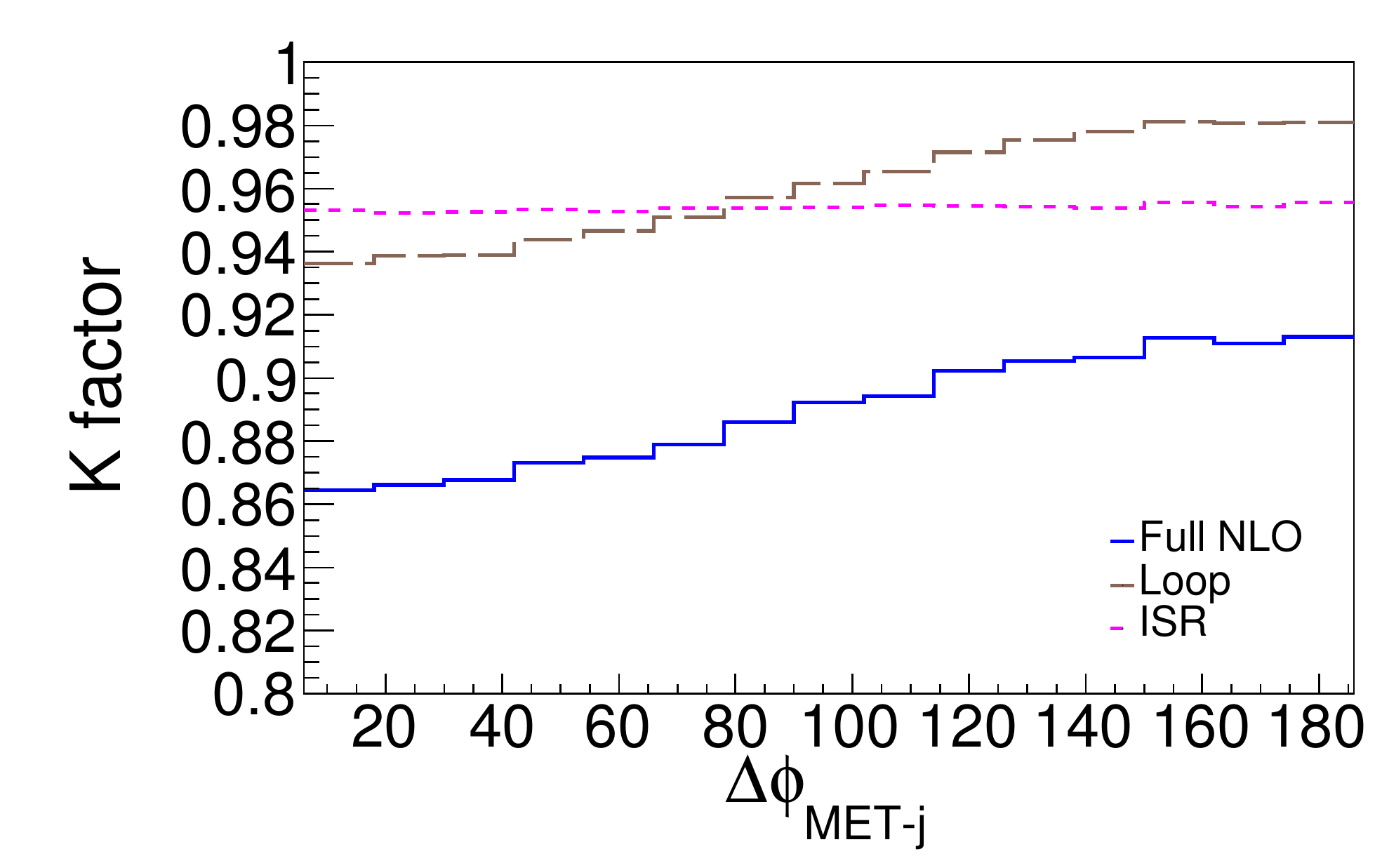}}
	\caption{Distribution in the azimuthal angle correlation between the tagging jet and the missing transverse energy (a), and the corresponding $K$ factors (b).}
\label{dphi}
\end{figure}
Finally, we present in Fig.~\ref{dphi} the distribution in the azimuthal angle between the tagging jet and missing transverse energy.
This distribution is sensitive to the anomalous $HWW$ couplings~\cite{Plehn:2001nj,Biswal:2012mp}, 
whereas in the SM the differential cross section exhibits only a steady increase with the increasing $\Delta\phi_{MET-j}$.
The relative EW corrections also display a mild dependence on $\Delta\phi_{MET-j}$ and change in the range between -14\% to -7\%. 
This shape distortion at NLO is caused by the loop contribution that is insensitive to the electron mass $m_e$. 
ISR in this case adds essentially a constant correction of -5\%. At this point, it is still hard to predict the role of the loop terms
in a systematic study of Higgs phenomenology at the LHeC.

\section{Conclusion}
\label{conclusion}
We have computed the NLO EW corrections for the Higgs production process via CC WBF at the LHeC. This is the first calculation that
takes into account the full EW effect at one loop for this process. To handle the singularities of various origins in the amplitudes,
we have worked in the dipole subtraction formalism and factorized the collinear radiation of the initial states. We have developed our own
program to implement the subtraction procedure and carry out the numerical integration over the phase space. Checks at many levels have been done
to verify the consistency of the calculation. 

For the CM energy at $1.98$ TeV, we have found a relative correction of 9\% (18\%) in the total cross section
with no cuts using the $G_{\mu}$ ($\alpha_{M_Z}$) renormalization scheme. The agreement between the two schemes are significantly improved when
the LO and NLO-correction terms are added. The differential cross sections for several observables have also been computed in the $G_{\mu}$ scheme 
under the selection cuts for the WBF process. The corrections for these variables are within -20\%. 
Sizable distortion of the distributions are also observed due to the ISR and loop corrections.  
Despite the smallness of the EW coupling $\alpha$ as compared with the strong coupling $\alpha_S$,
EW corrections are considerably larger than QCD corrections at NLO both because of 
the large number of diagrams (compared with the number of the QCD loop diagrams), 
and because of the ISR effect associated with photon radiations by the electron beam. 
The results show that the EW corrections is very important for the 
study of Higgs phenomenology at the LHeC and should not be neglected in a full analysis of the WBF process. 
\section{Acknowledgement}

The authors thank Lilin Yang, Ruibo Li, and Xiang Lv for helpful discussions. 
B.W. would like to thank Zhuoni Qian for useful suggestions on histograming, and Ted Rogers for providing a very important reference.
This work was supported by National Science Foundation of China (11875232, 12147103, 12105068). 	
 B.W. was also supported by Hangzhou Normal University Start-up Funds.

\appendix

\section{Expressions for the 4-particle final states}
\label{app:4-particle-final-states}

The specific form of the subtraction term for the quark- and anti-quark-initiated process
\begin{equation}
\begin{aligned}
    e^-(p_a)+q(p_b) \rightarrow \nu_e(p_1)+\gamma(p_2)+q'(p_3)+H(p_4) \nonumber
\end{aligned}
\end{equation}
reads
\begin{equation}
\begin{aligned}
&|\mathcal{M}_{sub}|^2=\mathcal{D}_{q'\gamma}^{e}+\mathcal{D}_{\gamma,q'}^{e}+\mathcal{D}_{\gamma}^{e,q}+\mathcal{D}_{q'\gamma}^{q}+\mathcal{D}_{\gamma,q'}^{q}+\mathcal{D}_{\gamma}^{q,e} \\ 
&\qquad \quad =-\frac{1}{2 p_3 p_2} \frac{1}{x_{32,a}}8 \pi \mu^{2\epsilon} \alpha\left[\frac{2}{2-x_{32,a}-z_{3a}}-1-z_{3a}-\epsilon(1-z_{3a})\right] \boldsymbol{Q}_{3a}^2|\mathcal{M}^q_0(x_{32,a}p_a+P_{3a};x_{32,a}p_a)|^2\\
&\qquad \qquad -\frac{1}{2 p_a p_2} \frac{1}{x_{a2,3}}8 \pi \mu^{2\epsilon} \alpha\left[\frac{2}{2-x_{a2,3}-z_{3a}}-1-x_{a2,3}-\epsilon(1-x_{a2,3})\right] \boldsymbol{Q}_{a3}^2|\mathcal{M}^q_0(x_{a2,3}p_a+P_{3a};x_{a2,3}p_a)|^2\\
&\qquad \qquad -\frac{1}{2 p_a p_2} \frac{1}{x_{a2,b}}8 \pi \mu^{2\epsilon} \alpha\left[\frac{2}{1-x_{a2,b}}-1-x_{a2,b}-\epsilon(1-x_{a2,b})\right] \boldsymbol{Q}_{ab}^2|\mathcal{M}^q_0(\tilde{p}_i(x_{a2,b});x_{a2,b}p_a)|^2\\
&\qquad \qquad -\frac{1}{2 p_3 p_2} \frac{1}{x_{32,b}} 8 \pi \mu^{2\epsilon} \alpha\left[\frac{2}{2-x_{32,b}-z_{3b}}-1-z_{3b}-\epsilon(1-z_{3b})\right] \boldsymbol{Q}_{3b}^2|\mathcal{M}^q_0(x_{32,b}p_b+P_{3b};x_{32,b}p_b)|^2 \\
&\qquad \qquad -\frac{1}{2 p_b p_2} \frac{1}{x_{b2,3}} 8 \pi \mu^{2\epsilon} \alpha\left[\frac{2}{2-x_{b2,3}-z_{3b}}-1-x_{b2,3}-\epsilon(1-x_{b2,3})\right] \boldsymbol{Q}_{b3}^2|\mathcal{M}^q_0(x_{b2,3}p_b+P_{3b};x_{b2,3}p_b)|^2\\
&\qquad \qquad -\frac{1}{2 p_b p_2} \frac{1}{x_{b2,a}} 8 \pi \mu^{2\epsilon} \alpha\left[\frac{2}{1-x_{b2,a}}-1-x_{b2,a}-\epsilon(1-x_{b2,a})\right] \boldsymbol{Q}_{ba}^2|\mathcal{M}^q_0(\tilde{p}_i(x_{b2,a});x_{b2,a}p_b)|^2,
\label{eq:realdipq}
\end{aligned}
\end{equation}
where $\mathcal{M}^q_0$ is the amplitude for the born process $e q \rightarrow \nu_e q' H$. The momentum fractions are defined as
\begin{equation}
\begin{aligned}
&x_{32,a}=x_{a2,3}=1-\frac{p_3p_2}{(p_3+p_2)p_a}, \qquad z_{3a}=\frac{p_3p_a}{(p_3+p_2)p_a},\\ &x_{32,b}=x_{b2,3}=1-\frac{p_3p_2}{(p_3+p_2)p_b}, \qquad z_{3b}=\frac{p_3p_b}{(p_3+p_2)p_b}\\
&x_{a2,b}=x_{b2,a}=1-\frac{p_2(p_a+p_b)}{p_ap_b}.
\label{eq:defx}
\end{aligned}
\end{equation}
As remarked in Sec.~\ref{4-part}, the born amplitudes in Eqs.~\ref{eq:realdipq} depend on the shifted momenta obtained from the 
corresponding radiative processes.
These momenta are shown as the arguments of the born amplitudes, where the momenta from the initial/final state is on the right/left
of the semicolon (Momenta with no modifications are not shown).
They are related to the momenta of the real emission processes by
\begin{equation}
\begin{aligned}
&P_{3a}=p_3+p_2-p_a=p_3+p_2-(1-x_{32,a})p_a-x_{32,a}p_a, \\
&P_{3b}=p_3+p_2-p_b=p_3+p_2-(1-x_{32,b})p_b-x_{32,b}p_b, \\
&P_{ab}=p_a+p_b-p_2, \qquad \tilde{P}_{ab}=x_{a2,b}p_a+p_b, \\
&\tilde{p}_i^{\mu}=\Lambda_{\nu}^{\mu}p_i^{\nu}=\bigg(g_{\nu}^{\mu}-\frac{(P_{ab}+\tilde{P}_{ab})^{\mu}(P_{ab}+\tilde{P}_{ab})_{\nu}}{P_{ab}^2+P_{ab}\tilde{P}_{ab}}+\frac{\tilde{P}_{ab}^{\mu}P_{ab,\nu}}{P_{ab}^2}\bigg)p_i^{\nu}.
\label{eq:Momtrans}
\end{aligned}
\end{equation}
Note that $\tilde{p}_i$ for all final state particles in $\mathcal{D}_{\gamma}^{e,q}$ and $\mathcal{D}_{\gamma}^{q,e}$ 
must transform according to Eq.~\ref{eq:Momtrans}.
The charge correlator $\boldsymbol{Q}_{ik}^2$ is defined by the charges of the flavor $i$ and $k$ as
\begin{equation}
\begin{aligned}
	&\boldsymbol{Q}_{ik}^2=Q_iQ_k\theta_i\theta_k,
\label{eq:defQ-gen}
\end{aligned}
\end{equation}
where $\theta_{i/k}$ is $1$ ($-1$) if it is in the final (initial) state.
For quark-initiated processes in our calculation, i.e., $q=u,\, c$, and $q'=d,\,s$, the charge correlators in Eq.~\ref{eq:realdipq} are
\begin{equation}
\begin{aligned}
	&\boldsymbol{Q}_{3a}^2=\boldsymbol{Q}_{a3}^2=Q_3(-Q_a)=-\frac{1}{3},\qquad \boldsymbol{Q}_{3b}^2=\boldsymbol{Q}_{b3}^2=Q_3(-Q_b)=\frac{2}{9},\\ &\boldsymbol{Q}_{ab}^2=\boldsymbol{Q}_{ba}^2=(-Q_a)(-Q_b)=-\frac{2}{3},
\label{eq:defQ-quark}
\end{aligned}
\end{equation}
where each initial state fermion acquires a minus sign on its charge.
The anti-quark-initiated processes, with $q=\bar{d},\, \bar{s}$, and $q'=\bar{u},\,\bar{c}$, differ from the corresponding quark-initiated processes only by the charge correlators, which are
\begin{equation}
\begin{aligned}
&\boldsymbol{Q}_{3a}^2=\boldsymbol{Q}_{a3}^2=-\frac{2}{3}, \qquad \boldsymbol{Q}_{3b}^2=\boldsymbol{Q}_{b3}^2=\frac{2}{9},\qquad \boldsymbol{Q}_{ab}^2=\boldsymbol{Q}_{ba}^2=-\frac{1}{3}.
\end{aligned}
\end{equation}

For the photon-induced processes
\begin{equation}
\begin{aligned}
    e^-(p_a)+\gamma(p_b) \rightarrow \nu_e(p_1)+q(p_2)+q'(p_3)+H(p_4) \nonumber
\end{aligned}
\end{equation}
the subtraction term takes the form
\begin{equation}
\begin{aligned}
|\mathcal{M}_{sub}|^2&=\mathcal{D}_{q,q'}^{\gamma}+\mathcal{D}_{q}^{\gamma,e}+\mathcal{D}_{q',q}^{\gamma}+\mathcal{D}_{q'}^{\gamma,e}\\
	&=-\frac{1}{2p_bp_2}\frac{1}{x_{b2,3}}8\pi \mu^{2\epsilon} \alpha \Big[1-\epsilon -2x_{b2,3}(1-x_{b2,3})\Big]N_{C,f}\boldsymbol{Q}_{\widetilde{b2}3}^2|\mathcal{M}^{\gamma}_0(x_{b2,3}p_b+P_{3b},x_{b2,3}p_b)|^2\\
&\quad -\frac{1}{2p_bp_2}\frac{1}{x_{b2,a}}8\pi \mu^{2\epsilon} \alpha \Big[1-\epsilon -2x_{b2,a}(1-x_{b2,a})\Big]N_{C,f}\boldsymbol{Q}_{\widetilde{b2}a}^2|\mathcal{M}^{\gamma}_0(\tilde{p}_i,x_{b2,a}p_b)|^2\\
&\quad -\frac{1}{2p_bp_3}\frac{1}{x_{b3,2}}8\pi \mu^{2\epsilon} \alpha \Big[1-\epsilon -2x_{b3,2}(1-x_{b3,2})\Big]N_{C,f}\boldsymbol{Q}_{\widetilde{b3}2}^2|\mathcal{M}^{\gamma}_0(x_{b3,2}p_b+P_{3b},x_{b3,2}p_b)|^2\\
&\quad -\frac{1}{2p_bp_3}\frac{1}{x_{b3,a}}8\pi \mu^{2\epsilon} \alpha \Big[1-\epsilon -2x_{b3,a}(1-x_{b3,a})\Big]N_{C,f}\boldsymbol{Q}_{\widetilde{b3}a}^2|\mathcal{M}^{\gamma}_0(\tilde{p}_i,x_{b3,a}p_b)|^2,
\end{aligned}
\end{equation}
where $\mathcal{M}^{\gamma}_0$ is the amplitude for the born process $e \bar{q} \rightarrow \nu_e q' H$ 
(or equivalently, $e \bar{q'} \rightarrow \nu_e q H$). In addition to the definitions made in Eq.~\ref{eq:defx}, we have
\begin{equation}
\begin{aligned}
	&x_{b3,2}=1-\frac{p_3p_2}{(p_3+p_2)p_b}, \qquad x_{b3,a}=1-\frac{p_3(p_a+p_b)}{p_ap_b}, \qquad N_{C,f}=3.
\end{aligned}
\end{equation}
The ``tilde'' symbol in the subscript of each charge correlator denotes the flavor 
from the photon splitting that enters the corresponding born process, 
namely, $\widetilde{b2}$ denotes $\bar{q}$ and $\widetilde{b3}$ denotes $\bar{q'}$.
The charge correlators for $q=\bar{u},\, \bar{c}$, and $q'=d,\,s$ are
\begin{equation}
\begin{aligned}
&\boldsymbol{Q}_{\widetilde{b2}3}^2=\frac{2}{9}, \qquad \boldsymbol{Q}_{\widetilde{b3}2}^2=\frac{2}{9}, \qquad \boldsymbol{Q}_{\widetilde{b2}a}^2=-\frac{2}{3}, \qquad \boldsymbol{Q}_{\widetilde{b3}a}^2=-\frac{1}{3}.
\end{aligned}
\end{equation}

\section{Expressions for the 3-particle final states}
\label{app:3-particle-final-states}

In this appendix we list the specific form of the $\boldsymbol{I}$,$\boldsymbol{K}$ and $\boldsymbol{P}$ terms 
that are factorized from the born process
\begin{equation}
\begin{aligned}
    e^-+q \rightarrow \nu_e(p_1)+q'(p_2)+H(p_3), \nonumber
\end{aligned}
\end{equation}
where the momenta of the initial particles are not shown explicitly.
For the corresponding $\boldsymbol{I}$,$\boldsymbol{K}$ and $\boldsymbol{P}$ terms, 
they are given by the arguments of the born factors in the first three lines of Eq.~\ref{eq:virtualsub}.
Recall that there $p_a=P_A$, and $p_b=\eta_bP_B$.

The generic expressions for the quark- and anti-quark-initiated processes are 
\begin{equation}
\begin{aligned}
\boldsymbol{I}^q(\epsilon)&=-\frac{\alpha}{2 \pi} \frac{(4 \pi)^{\epsilon}}{\Gamma(1-\epsilon)}\left[\frac{1}{\epsilon^{2}}+\frac{3}{2 \epsilon}+5-\frac{\pi^{2}}{2}\right] \Bigg\{\boldsymbol{Q}_{2a}^{2}\left(\frac{\mu^{2}}{2 p_{2} p_{a}}\right)^{\epsilon}+\boldsymbol{Q}_{a2}^{2}\left(\frac{\mu^{2}}{2 p_{a} p_{2}}\right)^{\epsilon}\\
&\qquad \qquad +\boldsymbol{Q}_{ab}^{2}\left(\frac{\mu^{2}}{2 p_{a} p_{b}}\right)^{\epsilon}+\boldsymbol{Q}_{2b}^{2}\left(\frac{\mu^{2}}{2 p_{2} p_{b}}\right)^{\epsilon}+\boldsymbol{Q}_{b2}^{2}\left(\frac{\mu^{2}}{2 p_{b} p_{2}}\right)^{\epsilon}+\boldsymbol{Q}_{ba}^{2}\left(\frac{\mu^{2}}{2 p_{b} p_{a}}\right)^{\epsilon}\Bigg\},
\end{aligned}
\end{equation}
\begin{equation}
\begin{aligned}
	\boldsymbol{K}^{c'}_{ff}(x)&=\frac{\alpha}{2\pi}\Bigg\{-Q_{f}^2(1+x)\log \frac{1-x}{x}+Q_{f}^2(1-x)+\bigg[Q_f^2\Big(\frac{2}{1-x}\log \frac{1-x}{x}\Big)_+ -\delta(1-x)\Big((5-\pi^2)Q_f^2\Big)\bigg]\\
	&\quad+\frac{3}{2}\frac{\boldsymbol{Q}_{2f}^2}{Q_f^2}\bigg[\Big(\frac{1}{1-x}\Big)_++\delta(1-x)\bigg]-\boldsymbol{Q}_{fc'}^2\Bigg[-(1+x)\log(1-x)+\bigg[2\bigg(\frac{\log(1-x)}{1-x}\bigg)_+ -\frac{\pi^2}{3}\delta(1-x)\bigg]\Bigg]\Bigg\},
\end{aligned}
\end{equation}
\begin{equation}
\begin{aligned}
	\boldsymbol{P}^{c'}_{ff}(x,\mu_F^2)&=\frac{\alpha}{2\pi}\bigg\{-(1+x)+2\Big(\frac{1}{1-x}\Big)_+ +\frac{3}{2}\delta(1-x)\bigg\}\bigg[\boldsymbol{Q}_{f2}^2\log \frac{\mu_F^2}{2xp_fp_2}+\boldsymbol{Q}_{f{c'}}^2\log \frac{\mu_F^2}{2xp_fp_{c'}}\bigg],
\end{aligned}
\end{equation}
where ``$q$'' in $\boldsymbol{I}^q$ can be a quark or anti-quark. In $\boldsymbol{K}^b_{ff}$ and  $\boldsymbol{P}^b_{ff}$, ``$f$''
can be a quark, anti-quark, or electron, while ``$c'$'' is the initial flavor not going through 
splitting (``$c$'' without prime is reserved for the charm quark, which is only a particular case of $c'$. See below). 
Inserting the charge correlators for specific flavors, we find
\begin{equation}
\begin{aligned}
\boldsymbol{I}^{u/c}(\epsilon)&=
&=-\frac{\alpha}{2 \pi} \frac{(4 \pi \mu^2)^{\epsilon}}{\Gamma(1-\epsilon)}\left[\frac{1}{\epsilon^{2}}+\frac{3}{2 \epsilon}+5-\frac{\pi^{2}}{2}\right] \Bigg\{-\frac{2}{3}\left(\frac{1}{2 p_{a} p_{2}}\right)^{\epsilon}-\frac{4}{3}\left(\frac{1}{2 p_{a} p_{b}}\right)^{\epsilon}+\frac{4}{9}\left(\frac{1}{2 p_{b} p_{2}}\right)^{\epsilon}\Bigg\},
\end{aligned}
\end{equation}
\begin{equation}
\begin{aligned}
\boldsymbol{I}^{\bar{d}/\bar{s}}(\epsilon)=-\frac{\alpha}{2 \pi} \frac{(4 \pi \mu^2)^{\epsilon}}{\Gamma(1-\epsilon)}\left[\frac{1}{\epsilon^{2}}+\frac{3}{2 \epsilon}+5-\frac{\pi^{2}}{2}\right] \Bigg\{-\frac{4}{3}\left(\frac{1}{2 p_{a} p_{2}}\right)^{\epsilon}-\frac{2}{3}\left(\frac{1}{2 p_{a} p_{b}}\right)^{\epsilon}+\frac{4}{9}\left(\frac{1}{2 p_{b} p_{2}}\right)^{\epsilon}\Bigg\},
\end{aligned}
\end{equation}
\begin{equation}
\begin{aligned}
\boldsymbol{K}^{u/c}_{ee}(x)&=\frac{\alpha}{2\pi}\bigg\{-(1+x)\log \frac{1-x}{x}+(1-x)+\Big(\frac{2}{1-x}\log \frac{1-x}{x}\Big)_+ -\frac{9}{2}\Big(\frac{1}{1-x}\Big)_+-\frac{2}{3}(1+x)\log(1-x)\\
&\qquad \qquad +\frac{4}{3}\Big(\frac{\log(1-x)}{1-x}\Big)_+ +\Big(\frac{7}{9}\pi^2-\frac{19}{2}\Big)\delta(1-x)\bigg\},
\end{aligned}
\end{equation}
\begin{equation}
\begin{aligned}
\boldsymbol{K}^e_{uu/cc}(x)&=\frac{\alpha}{2\pi}\bigg\{-\frac{4}{9}(1+x)\log\frac{1-x}{x}+\frac{4}{9}(1-x)+\frac{4}{9}\Big(\frac{2}{1-x}\log \frac{1-x}{x}\Big)_++3\Big(\frac{1}{1-x}\Big)_+-\frac{2}{3}(1+x)\log(1-x)\\
&\qquad \qquad +\frac{4}{3}\Big(\frac{\log(1-x)}{1-x}\Big)_++\Big(\frac{2}{9}\pi^2+\frac{7}{9}\Big)\delta(1-x)\bigg\},
\end{aligned}
\end{equation}
\begin{equation}
\begin{aligned}
	\boldsymbol{K}^{\bar{d}/\bar{s}}_{ee}(x)&=\frac{\alpha}{2\pi}\bigg\{-(1+x)\log \frac{1-x}{x}+(1-x)+\Big(\frac{2}{1-x}\log \frac{1-x}{x}\Big)_+-\frac{9}{4}\Big(\frac{1}{1-x}\Big)_+-\frac{1}{3}(1+x)\log(1-x) \\
&\qquad \qquad +\frac{2}{3}\Big(\frac{\log(1-x)}{1-x}\Big)_++\Big(\frac{8}{9}\pi^2-\frac{29}{4}\Big)\delta(1-x)\bigg\},
\end{aligned}
\end{equation}
\begin{equation}
\begin{aligned}
\boldsymbol{K}^e_{\bar{d}\bar{d}/\bar{s}\bar{s}}(x)&=\frac{\alpha}{2\pi}\bigg\{-\frac{1}{9}(1+x)\log \frac{1-x}{x}+\frac{1}{9}(1-x)+\frac{1}{9}\Big(\frac{2}{1-x}\log\frac{1-x}{x}\Big)_+ +\frac{3}{4}\Big(\frac{1}{1-x}\Big)_+-\frac{1}{3}(1+x)\log(1-x)\\
&\qquad \qquad +\frac{2}{3}\bigg(\frac{\log(1-x)}{1-x}\bigg)_++\frac{7}{36}\delta(1-x)\bigg\},
\end{aligned}
\end{equation}
\begin{equation}
\begin{aligned}
\boldsymbol{P}^{u/c}_{ee}(x,\mu_F^2)=\frac{\alpha}{2\pi}\bigg\{-(1+x)+2\Big(\frac{1}{1-x}\Big)_+ +\frac{3}{2}\delta(1-x)\bigg\}\bigg[-\frac{1}{3}\log \frac{\mu_F^2}{2xp_ap_2}-\frac{2}{3}\log \frac{\mu_F^2}{2xp_ap_b}\bigg],
\end{aligned}
\end{equation}
\begin{equation}
\begin{aligned}
\boldsymbol{P}^e_{uu/cc}(x,\mu_F^2)=\frac{\alpha}{2\pi}\bigg\{-(1+x)+2\Big(\frac{1}{1-x}\Big)_+ +\frac{3}{2}\delta(1-x)\bigg\}\bigg[\frac{2}{9}\log \frac{\mu_F^2}{2xp_bp_2}-\frac{2}{3}\log \frac{\mu_F^2}{2xp_ap_b}\bigg],
\end{aligned}
\end{equation}
\begin{equation}
\begin{aligned}
\boldsymbol{P}^{\bar{d}/\bar{s}}_{ee}(x,\mu_F^2)=\frac{\alpha}{2\pi}\bigg\{-(1+x)+2\Big(\frac{1}{1-x}\Big)_+ +\frac{3}{2}\delta(1-x)\bigg\}\bigg[-\frac{2}{3}\log \frac{\mu_F^2}{2xp_ap_2}-\frac{1}{3}\log \frac{\mu_F^2}{2xp_ap_b}\bigg],
\end{aligned}
\end{equation}
\begin{equation}
\begin{aligned}
\boldsymbol{P}^e_{\bar{d}\bar{d}/\bar{s}\bar{s}}(x,\mu_F^2)=\frac{\alpha}{2\pi}\bigg\{-(1+x)+2\Big(\frac{1}{1-x}\Big)_+ +\frac{3}{2}\delta(1-x)\bigg\}\bigg[\frac{2}{9}\log \frac{\mu_F^2}{2xp_bp_2}-\frac{1}{3}\log \frac{\mu_F^2}{2xp_ap_b}\bigg].
\end{aligned}
\end{equation}

For the photon-initiated processes, $\boldsymbol{I}$ terms do not contribute. 
Also the electron splitting terms are of higher order in $\alpha$ and do not contribute in this case either. $\boldsymbol{K}$ and $\boldsymbol{P}$ terms take the form
\begin{equation}
\begin{aligned}
	\boldsymbol{K}^a_{\gamma f}(x)&=\frac{\alpha}{2\pi}\Bigg\{N_{C,f}Q_f^2\frac{1+(1-x)^2}{x}\log\frac{1-x}{x}+N_{C,f}Q_f^2x(1-x)-\boldsymbol{Q}_{fa}^2N_{C,f}\frac{1+(1-x)^2}{x}\log(1-x)\Bigg\},
\end{aligned}
\end{equation}
\begin{equation}
\begin{aligned}
	\boldsymbol{P}^a_{\gamma f}(x,\mu_F^2)=\frac{\alpha}{2\pi}\bigg\{N_{C,f}\frac{1+(1-x)^2}{x}\bigg\}\bigg[\boldsymbol{Q}_{f2}^2\log \frac{\mu_F^2}{2xp_bp_2}+\boldsymbol{Q}_{fa}^2\log \frac{\mu_F^2}{2xp_ap_b}\bigg].
\end{aligned}
\end{equation}
The expressions for specific splitting processes are
\begin{equation}
\begin{aligned}
\boldsymbol{K}^e_{\gamma u/c}(x)=\frac{\alpha}{2\pi}\Bigg\{\frac{4}{3}\frac{1+(1-x)^2}{x}\log\frac{1-x}{x}+\frac{4}{3}x(1-x)+2\frac{1+(1-x)^2}{x}\log(1-x)\Bigg\},
\end{aligned}
\end{equation}
\begin{equation}
\begin{aligned}
\boldsymbol{K}^e_{\gamma \bar{d}/\bar{s}}(x)=\frac{\alpha}{2\pi}\Bigg\{\frac{1}{3}\frac{1+(1-x)^2}{x}\log\frac{1-x}{x}+\frac{1}{3}x(1-x)+\frac{1+(1-x)^2}{x}\log(1-x)\Bigg\},
\end{aligned}
\end{equation}
\begin{equation}
\begin{aligned}
\boldsymbol{P}^e_{\gamma u/c}(x,\mu_F^2)=\frac{\alpha}{2\pi}\bigg\{\frac{4}{3}\frac{1+(1-x)^2}{x}\bigg\}\bigg[\frac{1}{2}\log \frac{\mu_F^2}{2xp_bp_2}-\frac{3}{2}\log \frac{\mu_F^2}{2xp_ap_b}\bigg],
\end{aligned}
\end{equation}
\begin{equation}
\begin{aligned}
\boldsymbol{P}^e_{\gamma \bar{d}/\bar{s}}(x,\mu_F^2)=\frac{\alpha}{2\pi}\bigg\{\frac{1}{3}\frac{1+(1-x)^2}{x}\bigg\}\bigg[2\log \frac{\mu_F^2}{2xp_bp_2}-3\log \frac{\mu_F^2}{2xp_ap_b}\bigg].
\end{aligned}
\end{equation}

\bibliographystyle{utphysmcite}
\bibliography{WBF}

\end{document}